\pdfoutput=1
\documentclass[9pt,a4paper,twocolumn]{article}

\usepackage[utf8]{inputenc}

\usepackage{libertine}
\usepackage[libertine]{newtxmath}

\usepackage{amssymb}
\usepackage{graphicx}
\usepackage[pdfa,pdfproducer={},pdfcreator={}]{hyperref}
\usepackage{url}

\usepackage{xcolor}
\usepackage{colortbl} 

\definecolor{bleudefrance}{rgb}{0.0, 0.28, 0.67}
\definecolor{bleudefrance}{rgb}{0., 0.37, 0.67}

\definecolor{histoblue}{HTML}{3981bf}
\definecolor{histobrown}{HTML}{99886e}
\definecolor{histoyellow}{HTML}{c2c099}

\definecolor{verdevida}{RGB}{108,193,76}
\definecolor{verdevida2}{RGB}{56,100,39}
\hypersetup{
	colorlinks = true,
	citecolor ={bleudefrance},
	allcolors= {bleudefrance},
	linkbordercolor = {white},
}

\usepackage{textgreek}

\usepackage[font=small,justification=RaggedRight]{caption}

\usepackage{pgfgantt}

\usepackage{microtype}
\usepackage[small,compact]{titlesec}

\titleformat{\section}{\bfseries\sffamily\scshape\color{black}}{\arabic{section}}{1em}{\centering\MakeUppercase}

\titleformat{\subsection}{\raggedright\bfseries\sffamily\scshape\small}{\arabic{section}.\arabic{subsection}}{1em}{\MakeUppercase}
\titleformat{\subsubsection}{\centering\bfseries\sffamily\scshape\footnotesize}{\arabic{section}.\arabic{subsection}.\arabic{subsubsection}}{1em}{\MakeUppercase}

\titlespacing{\subsubsection}{0pt}{*4}{*1}
\titlespacing{\subsection}{0pt}{*4}{*1}
\titlespacing{\section}{0pt}{*5}{*2}

\usepackage{fancyheadings}
\usepackage[left=1.50cm, right=1.50cm, top=1.5cm, bottom=2.00cm]{geometry}
\usepackage[backend=bibtex,isbn=true,url=false,eprint=false, bibstyle=numeric,sorting=none,    style=phys,articletitle=false,biblabel=superscript]{biblatex}

\AtEveryBibitem{%
	\clearfield{note}%
}

\DeclareCiteCommand{\citenum}
{}
{\printfield{labelnumber}}
{}
{}

\def\blx@bblfile@bibtex{
	\blx@secinit
	\begingroup
	\blx@bblstart
\begingroup
\makeatletter
\@ifundefined{ver@biblatex.sty}
  {\@latex@error
     {Missing 'biblatex' package}
     {The bibliography requires the 'biblatex' package.}
      \aftergroup }
  {}
\endgroup

\sortlist[entry]{none/global/}
  \entry{feller2008survey}{article}{}
    \name{author}{3}{}{%
      {{hash=FD}{%
         family={Feller},
         familyi={F\bibinitperiod},
         given={David},
         giveni={D\bibinitperiod},
      }}%
      {{hash=PKA}{%
         family={Peterson},
         familyi={P\bibinitperiod},
         given={Kirk\bibnamedelima A},
         giveni={K\bibinitperiod\bibinitdelim A\bibinitperiod},
      }}%
      {{hash=DDA}{%
         family={Dixon},
         familyi={D\bibinitperiod},
         given={David\bibnamedelima A},
         giveni={D\bibinitperiod\bibinitdelim A\bibinitperiod},
      }}%
    }
    \strng{namehash}{FD+1}
    \strng{fullhash}{FDPKADDA1}
    \field{labelnamesource}{author}
    \field{labeltitlesource}{title}
    \verb{doi}
    \verb 10.1063/1.3008061
    \endverb
    \field{number}{20}
    \field{pages}{204105}
    \field{title}{A survey of factors contributing to accurate theoretical
  predictions of atomization energies and molecular structures}
    \field{volume}{129}
    \field{journaltitle}{J. Chem. Phys.}
    \field{year}{2008}
  \endentry

  \entry{RAGHAVACHARI1989479}{article}{}
    \name{author}{4}{}{%
      {{hash=RK}{%
         family={Raghavachari},
         familyi={R\bibinitperiod},
         given={Krishnan},
         giveni={K\bibinitperiod},
      }}%
      {{hash=TGW}{%
         family={Trucks},
         familyi={T\bibinitperiod},
         given={Gary\bibnamedelima W.},
         giveni={G\bibinitperiod\bibinitdelim W\bibinitperiod},
      }}%
      {{hash=PJA}{%
         family={Pople},
         familyi={P\bibinitperiod},
         given={John\bibnamedelima A.},
         giveni={J\bibinitperiod\bibinitdelim A\bibinitperiod},
      }}%
      {{hash=HGM}{%
         family={Head-Gordon},
         familyi={H\bibinitperiod-G\bibinitperiod},
         given={Martin},
         giveni={M\bibinitperiod},
      }}%
    }
    \strng{namehash}{RK+1}
    \strng{fullhash}{RKTGWPJAHGM1}
    \field{labelnamesource}{author}
    \field{labeltitlesource}{title}
    \verb{doi}
    \verb https://doi.org/10.1016/S0009-2614(89)87395-6
    \endverb
    \field{issn}{0009-2614}
    \field{number}{6}
    \field{pages}{479 \bibrangedash  483}
    \field{title}{A fifth-order perturbation comparison of electron correlation
  theories}
    \verb{url}
    \verb http://www.sciencedirect.com/science/article/pii/S0009261489873956
    \endverb
    \field{volume}{157}
    \field{journaltitle}{Chem. Phys. Lett.}
    \field{year}{1989}
  \endentry

  \entry{ccsd_t_original}{article}{}
    \name{author}{3}{}{%
      {{hash=WJD}{%
         family={Watts},
         familyi={W\bibinitperiod},
         given={John\bibnamedelima D.},
         giveni={J\bibinitperiod\bibinitdelim D\bibinitperiod},
      }}%
      {{hash=GJ}{%
         family={Gauss},
         familyi={G\bibinitperiod},
         given={Jürgen},
         giveni={J\bibinitperiod},
      }}%
      {{hash=BRJ}{%
         family={Bartlett},
         familyi={B\bibinitperiod},
         given={Rodney\bibnamedelima J.},
         giveni={R\bibinitperiod\bibinitdelim J\bibinitperiod},
      }}%
    }
    \strng{namehash}{WJD+1}
    \strng{fullhash}{WJDGJBRJ1}
    \field{labelnamesource}{author}
    \field{labeltitlesource}{title}
    \verb{doi}
    \verb 10.1063/1.464480
    \endverb
    \verb{eprint}
    \verb https://doi.org/10.1063/1.464480
    \endverb
    \field{number}{11}
    \field{pages}{8718\bibrangedash 8733}
    \field{title}{Coupled‐cluster methods with noniterative triple
  excitations for restricted open‐shell Hartree–Fock and other general
  single determinant reference functions. Energies and analytical gradients}
    \verb{url}
    \verb https://doi.org/10.1063/1.464480
    \endverb
    \field{volume}{98}
    \field{journaltitle}{J. Chem. Phys}
    \field{year}{1993}
  \endentry

  \entry{karton2011w4}{article}{}
    \name{author}{3}{}{%
      {{hash=KA}{%
         family={Karton},
         familyi={K\bibinitperiod},
         given={Amir},
         giveni={A\bibinitperiod},
      }}%
      {{hash=DS}{%
         family={Daon},
         familyi={D\bibinitperiod},
         given={Shauli},
         giveni={S\bibinitperiod},
      }}%
      {{hash=MJM}{%
         family={Martin},
         familyi={M\bibinitperiod},
         given={Jan\bibnamedelima ML},
         giveni={J\bibinitperiod\bibinitdelim M\bibinitperiod},
      }}%
    }
    \list{publisher}{1}{%
      {Elsevier}%
    }
    \strng{namehash}{KA+1}
    \strng{fullhash}{KADSMJM1}
    \field{labelnamesource}{author}
    \field{labeltitlesource}{title}
    \verb{doi}
    \verb 10.1016/j.cplett.2011.05.007
    \endverb
    \field{number}{4}
    \field{pages}{165\bibrangedash 178}
    \field{title}{W4-11: A high-confidence benchmark dataset for computational
  thermochemistry derived from first-principles W4 data}
    \field{volume}{510}
    \field{journaltitle}{Chem. Phys. Lett.}
    \field{year}{2011}
  \endentry

  \entry{feller2015improved}{inbook}{}
    \name{author}{3}{}{%
      {{hash=FD}{%
         family={Feller},
         familyi={F\bibinitperiod},
         given={David},
         giveni={D\bibinitperiod},
      }}%
      {{hash=PKA}{%
         family={Peterson},
         familyi={P\bibinitperiod},
         given={Kirk\bibnamedelima A.},
         giveni={K\bibinitperiod\bibinitdelim A\bibinitperiod},
      }}%
      {{hash=RB}{%
         family={Ruscic},
         familyi={R\bibinitperiod},
         given={Branko},
         giveni={B\bibinitperiod},
      }}%
    }
    \name{editor}{3}{}{%
      {{hash=WAK}{%
         family={Wilson},
         familyi={W\bibinitperiod},
         given={Angela\bibnamedelima K.},
         giveni={A\bibinitperiod\bibinitdelim K\bibinitperiod},
      }}%
      {{hash=PKA}{%
         family={Peterson},
         familyi={P\bibinitperiod},
         given={Kirk\bibnamedelima A.},
         giveni={K\bibinitperiod\bibinitdelim A\bibinitperiod},
      }}%
      {{hash=WDE}{%
         family={Woon},
         familyi={W\bibinitperiod},
         given={David\bibnamedelima E.},
         giveni={D\bibinitperiod\bibinitdelim E\bibinitperiod},
      }}%
    }
    \list{publisher}{1}{%
      {Springer Berlin Heidelberg}%
    }
    \strng{namehash}{FD+1}
    \strng{fullhash}{FDPKARB1}
    \field{labelnamesource}{author}
    \field{labeltitlesource}{title}
    \field{booktitle}{Thom H. Dunning, Jr.: A Festschrift from Theoretical
  Chemistry Accounts}
    \verb{doi}
    \verb 10.1007/978-3-662-47051-0_4
    \endverb
    \field{isbn}{978-3-662-47051-0}
    \field{pages}{31\bibrangedash 46}
    \field{title}{Improved accuracy benchmarks of small molecules using
  correlation consistent basis sets}
    \verb{url}
    \verb https://doi.org/10.1007/978-3-662-47051-0_4
    \endverb
    \list{location}{1}{%
      {Berlin, Heidelberg}%
    }
    \field{year}{2015}
  \endentry

  \entry{martin2000thermochemical}{article}{}
    \name{author}{4}{}{%
      {{hash=MJML}{%
         family={Martin},
         familyi={M\bibinitperiod},
         given={Jan M.\bibnamedelima L.},
         giveni={J\bibinitperiod\bibinitdelim M\bibinitperiod\bibinitdelim
  L\bibinitperiod},
      }}%
      {{hash=SA}{%
         family={Sundermann},
         familyi={S\bibinitperiod},
         given={Andreas},
         giveni={A\bibinitperiod},
      }}%
      {{hash=FPL}{%
         family={Fast},
         familyi={F\bibinitperiod},
         given={Patton\bibnamedelima L.},
         giveni={P\bibinitperiod\bibinitdelim L\bibinitperiod},
      }}%
      {{hash=TDG}{%
         family={Truhlar},
         familyi={T\bibinitperiod},
         given={Donald\bibnamedelima G.},
         giveni={D\bibinitperiod\bibinitdelim G\bibinitperiod},
      }}%
    }
    \strng{namehash}{MJML+1}
    \strng{fullhash}{MJMLSAFPLTDG1}
    \field{labelnamesource}{author}
    \field{labeltitlesource}{title}
    \verb{doi}
    \verb 10.1063/1.481960
    \endverb
    \verb{eprint}
    \verb https://doi.org/10.1063/1.481960
    \endverb
    \field{number}{4}
    \field{pages}{1348\bibrangedash 1358}
    \field{title}{Thermochemical analysis of core correlation and scalar
  relativistic effects on molecular atomization energies}
    \verb{url}
    \verb https://doi.org/10.1063/1.481960
    \endverb
    \field{volume}{113}
    \field{journaltitle}{J. Chem. Phys.}
    \field{year}{2000}
  \endentry

  \entry{moller1934note}{article}{}
    \name{author}{2}{}{%
      {{hash=MC}{%
         family={M{\o}ller},
         familyi={M\bibinitperiod},
         given={Chr},
         giveni={C\bibinitperiod},
      }}%
      {{hash=PMS}{%
         family={Plesset},
         familyi={P\bibinitperiod},
         given={Milton\bibnamedelima S},
         giveni={M\bibinitperiod\bibinitdelim S\bibinitperiod},
      }}%
    }
    \list{publisher}{1}{%
      {APS}%
    }
    \strng{namehash}{MCPMS1}
    \strng{fullhash}{MCPMS1}
    \field{labelnamesource}{author}
    \field{labeltitlesource}{title}
    \field{number}{7}
    \field{pages}{618}
    \field{title}{Note on an approximation treatment for many-electron systems}
    \field{volume}{46}
    \field{journaltitle}{Phys. Rev.}
    \field{year}{1934}
  \endentry

  \entry{cepa_ref0}{article}{}
    \name{author}{1}{}{%
      {{hash=MW}{%
         family={Meyer},
         familyi={M\bibinitperiod},
         given={Wilfried},
         giveni={W\bibinitperiod},
      }}%
    }
    \strng{namehash}{MW1}
    \strng{fullhash}{MW1}
    \field{labelnamesource}{author}
    \field{labeltitlesource}{title}
    \verb{doi}
    \verb 10.1002/qua.560050839
    \endverb
    \verb{eprint}
    \verb https://onlinelibrary.wiley.com/doi/pdf/10.1002/qua.560050839
    \endverb
    \field{number}{S5}
    \field{pages}{341\bibrangedash 348}
    \field{title}{Ionization energies of water from PNO-CI calculations}
    \verb{url}
    \verb https://onlinelibrary.wiley.com/doi/abs/10.1002/qua.560050839
    \endverb
    \field{volume}{5}
    \field{journaltitle}{Int. J. Quantum Chem.}
    \field{year}{1971}
  \endentry

  \entry{cepa_ref1}{article}{}
    \name{author}{1}{}{%
      {{hash=MW}{%
         family={Meyer},
         familyi={M\bibinitperiod},
         given={Wilfried},
         giveni={W\bibinitperiod},
      }}%
    }
    \strng{namehash}{MW1}
    \strng{fullhash}{MW1}
    \field{labelnamesource}{author}
    \field{labeltitlesource}{title}
    \verb{doi}
    \verb 10.1063/1.1679283
    \endverb
    \verb{eprint}
    \verb https://doi.org/10.1063/1.1679283
    \endverb
    \field{number}{3}
    \field{pages}{1017\bibrangedash 1035}
    \field{title}{PNO–CI Studies of electron correlation effects. I.
  Configuration expansion by means of nonorthogonal orbitals, and application
  to the ground state and ionized states of methane}
    \verb{url}
    \verb https://doi.org/10.1063/1.1679283
    \endverb
    \field{volume}{58}
    \field{journaltitle}{J. Chem. Phys.}
    \field{year}{1973}
  \endentry

  \entry{cepa_ref2}{article}{}
    \name{author}{1}{}{%
      {{hash=MW}{%
         family={Meyer},
         familyi={M\bibinitperiod},
         given={Wilfried},
         giveni={W\bibinitperiod},
      }}%
    }
    \strng{namehash}{MW1}
    \strng{fullhash}{MW1}
    \field{labelnamesource}{author}
    \field{labeltitlesource}{title}
    \field{abstract}{%
    Ab initio calculations of the potential curves of low laying electronic
  states of OH are performed on the basis of a variational configuration
  interaction wavefunction (PNO-CI) and the coupled electron pair approximation
  (CEPA).%
    }
    \verb{doi}
    \verb 10.1007/BF00548478
    \endverb
    \field{issn}{1432-2234}
    \field{number}{4}
    \field{pages}{277\bibrangedash 292}
    \field{title}{PNO-CI and CEPA studies of electron correlation effects}
    \verb{url}
    \verb https://doi.org/10.1007/BF00548478
    \endverb
    \field{volume}{35}
    \field{journaltitle}{Theor. Chim. Acta}
    \field{year}{1974}
    \warn{\item Invalid format of field 'month'}
  \endentry

  \entry{cepa_ref3}{article}{}
    \name{author}{2}{}{%
      {{hash=MW}{%
         family={Meyer},
         familyi={M\bibinitperiod},
         given={W.},
         giveni={W\bibinitperiod},
      }}%
      {{hash=RP}{%
         family={Rosmus},
         familyi={R\bibinitperiod},
         given={P.},
         giveni={P\bibinitperiod},
      }}%
    }
    \strng{namehash}{MWRP1}
    \strng{fullhash}{MWRP1}
    \field{labelnamesource}{author}
    \field{labeltitlesource}{title}
    \verb{doi}
    \verb 10.1063/1.431665
    \endverb
    \verb{eprint}
    \verb https://doi.org/10.1063/1.431665
    \endverb
    \field{number}{6}
    \field{pages}{2356\bibrangedash 2375}
    \field{title}{PNO–CI and CEPA studies of electron correlation effects.
  III. Spectroscopic constants and dipole moment functions for the ground
  states of the first‐row and second‐row diatomic hydrides}
    \verb{url}
    \verb https://doi.org/10.1063/1.431665
    \endverb
    \field{volume}{63}
    \field{journaltitle}{J. Chem. Phys.}
    \field{year}{1975}
  \endentry

  \entry{cepa_ref4}{article}{}
    \name{author}{4}{}{%
      {{hash=AR}{%
         family={Ahlrichs},
         familyi={A\bibinitperiod},
         given={R.},
         giveni={R\bibinitperiod},
      }}%
      {{hash=LH}{%
         family={Lischka},
         familyi={L\bibinitperiod},
         given={H.},
         giveni={H\bibinitperiod},
      }}%
      {{hash=SV}{%
         family={Staemmler},
         familyi={S\bibinitperiod},
         given={V.},
         giveni={V\bibinitperiod},
      }}%
      {{hash=KW}{%
         family={Kutzelnigg},
         familyi={K\bibinitperiod},
         given={W.},
         giveni={W\bibinitperiod},
      }}%
    }
    \strng{namehash}{AR+1}
    \strng{fullhash}{ARLHSVKW1}
    \field{labelnamesource}{author}
    \field{labeltitlesource}{title}
    \verb{doi}
    \verb 10.1063/1.430637
    \endverb
    \verb{eprint}
    \verb https://doi.org/10.1063/1.430637
    \endverb
    \field{number}{4}
    \field{pages}{1225\bibrangedash 1234}
    \field{title}{PNO–CI (pair natural orbital configuration interaction) and
  CEPA–PNO (coupled electron pair approximation with pair natural orbitals)
  calculations of molecular systems. I. Outline of the method for
  closed‐shell states}
    \verb{url}
    \verb https://doi.org/10.1063/1.430637
    \endverb
    \field{volume}{62}
    \field{journaltitle}{J. Chem. Phys.}
    \field{year}{1975}
  \endentry

  \entry{cepa_ref5}{article}{}
    \name{author}{2}{}{%
      {{hash=AR}{%
         family={Ahlrichs},
         familyi={A\bibinitperiod},
         given={Reinhart},
         giveni={R\bibinitperiod},
      }}%
      {{hash=DF}{%
         family={Driessler},
         familyi={D\bibinitperiod},
         given={Frank},
         giveni={F\bibinitperiod},
      }}%
    }
    \strng{namehash}{ARDF1}
    \strng{fullhash}{ARDF1}
    \field{labelnamesource}{author}
    \field{labeltitlesource}{title}
    \verb{doi}
    \verb 10.1007/BF00549691
    \endverb
    \field{issn}{1432-2234}
    \field{number}{4}
    \field{pages}{275\bibrangedash 287}
    \field{title}{Direct determination of pair natural orbitals}
    \verb{url}
    \verb https://doi.org/10.1007/BF00549691
    \endverb
    \field{volume}{36}
    \field{journaltitle}{Theor. Chim. Acta}
    \field{year}{1975}
    \warn{\item Invalid format of field 'month'}
  \endentry

  \entry{martin1999towards}{article}{}
    \name{author}{2}{}{%
      {{hash=MJM}{%
         family={Martin},
         familyi={M\bibinitperiod},
         given={Jan\bibnamedelima ML},
         giveni={J\bibinitperiod\bibinitdelim M\bibinitperiod},
      }}%
      {{hash=dOG}{%
         prefix={de},
         prefixi={d\bibinitperiod},
         family={Oliveira},
         familyi={O\bibinitperiod},
         given={Gl{\^e}nisson},
         giveni={G\bibinitperiod},
      }}%
    }
    \list{publisher}{1}{%
      {AIP Publishing}%
    }
    \strng{namehash}{MJMOGd1}
    \strng{fullhash}{MJMOGd1}
    \field{labelnamesource}{author}
    \field{labeltitlesource}{title}
    \verb{doi}
    \verb 10.1063/1.479454
    \endverb
    \field{number}{5}
    \field{pages}{1843\bibrangedash 1856}
    \field{title}{Towards standard methods for benchmark quality ab initio
  thermochemistry—W1 and W2 theory}
    \field{volume}{111}
    \field{journaltitle}{J. Chem. Phys.}
    \field{year}{1999}
  \endentry

  \entry{curtiss1998gaussian}{article}{}
    \name{author}{5}{}{%
      {{hash=CLA}{%
         family={Curtiss},
         familyi={C\bibinitperiod},
         given={Larry\bibnamedelima A},
         giveni={L\bibinitperiod\bibinitdelim A\bibinitperiod},
      }}%
      {{hash=RK}{%
         family={Raghavachari},
         familyi={R\bibinitperiod},
         given={Krishnan},
         giveni={K\bibinitperiod},
      }}%
      {{hash=RPC}{%
         family={Redfern},
         familyi={R\bibinitperiod},
         given={Paul\bibnamedelima C},
         giveni={P\bibinitperiod\bibinitdelim C\bibinitperiod},
      }}%
      {{hash=RV}{%
         family={Rassolov},
         familyi={R\bibinitperiod},
         given={Vitaly},
         giveni={V\bibinitperiod},
      }}%
      {{hash=PJA}{%
         family={Pople},
         familyi={P\bibinitperiod},
         given={John\bibnamedelima A},
         giveni={J\bibinitperiod\bibinitdelim A\bibinitperiod},
      }}%
    }
    \list{publisher}{1}{%
      {AIP Publishing}%
    }
    \strng{namehash}{CLA+1}
    \strng{fullhash}{CLARKRPCRVPJA1}
    \field{labelnamesource}{author}
    \field{labeltitlesource}{title}
    \verb{doi}
    \verb 10.1063/1.477422
    \endverb
    \field{number}{18}
    \field{pages}{7764\bibrangedash 7776}
    \field{title}{Gaussian-3 (G3) theory for molecules containing first and
  second-row atoms}
    \field{volume}{109}
    \field{journaltitle}{J. Chem. Phys.}
    \field{year}{1998}
  \endentry

  \entry{curtiss2007gaussian}{article}{}
    \name{author}{3}{}{%
      {{hash=CLA}{%
         family={Curtiss},
         familyi={C\bibinitperiod},
         given={Larry\bibnamedelima A},
         giveni={L\bibinitperiod\bibinitdelim A\bibinitperiod},
      }}%
      {{hash=RPC}{%
         family={Redfern},
         familyi={R\bibinitperiod},
         given={Paul\bibnamedelima C},
         giveni={P\bibinitperiod\bibinitdelim C\bibinitperiod},
      }}%
      {{hash=RK}{%
         family={Raghavachari},
         familyi={R\bibinitperiod},
         given={Krishnan},
         giveni={K\bibinitperiod},
      }}%
    }
    \list{publisher}{1}{%
      {AIP Publishing}%
    }
    \strng{namehash}{CLA+1}
    \strng{fullhash}{CLARPCRK1}
    \field{labelnamesource}{author}
    \field{labeltitlesource}{title}
    \verb{doi}
    \verb 10.1063/1.2436888
    \endverb
    \field{number}{8}
    \field{pages}{084108}
    \field{title}{Gaussian-4 theory}
    \field{volume}{126}
    \field{journaltitle}{J. Chem. Phys.}
    \field{year}{2007}
  \endentry

  \entry{doser2008tighter}{article}{}
    \name{author}{3}{}{%
      {{hash=DB}{%
         family={Doser},
         familyi={D\bibinitperiod},
         given={Bernd},
         giveni={B\bibinitperiod},
      }}%
      {{hash=LDS}{%
         family={Lambrecht},
         familyi={L\bibinitperiod},
         given={Daniel\bibnamedelima S},
         giveni={D\bibinitperiod\bibinitdelim S\bibinitperiod},
      }}%
      {{hash=OC}{%
         family={Ochsenfeld},
         familyi={O\bibinitperiod},
         given={Christian},
         giveni={C\bibinitperiod},
      }}%
    }
    \list{publisher}{1}{%
      {Royal Society of Chemistry}%
    }
    \strng{namehash}{DB+1}
    \strng{fullhash}{DBLDSOC1}
    \field{labelnamesource}{author}
    \field{labeltitlesource}{title}
    \verb{doi}
    \verb 10.1039/B804110E
    \endverb
    \field{number}{23}
    \field{pages}{3335\bibrangedash 3344}
    \field{title}{Tighter multipole-based integral estimates and parallel
  implementation of linear-scaling AO--MP2 theory}
    \field{volume}{10}
    \field{journaltitle}{Phys. Chem. Chem. Phys.}
    \field{year}{2008}
  \endentry

  \entry{steele2006dual}{article}{}
    \name{author}{5}{}{%
      {{hash=SRP}{%
         family={Steele},
         familyi={S\bibinitperiod},
         given={Ryan\bibnamedelima P},
         giveni={R\bibinitperiod\bibinitdelim P\bibinitperiod},
      }}%
      {{hash=DJRA}{%
         family={DiStasio\bibnamedelima Jr},
         familyi={D\bibinitperiod\bibinitdelim J\bibinitperiod},
         given={Robert\bibnamedelima A},
         giveni={R\bibinitperiod\bibinitdelim A\bibinitperiod},
      }}%
      {{hash=SY}{%
         family={Shao},
         familyi={S\bibinitperiod},
         given={Yihan},
         giveni={Y\bibinitperiod},
      }}%
      {{hash=KJ}{%
         family={Kong},
         familyi={K\bibinitperiod},
         given={Jing},
         giveni={J\bibinitperiod},
      }}%
      {{hash=HGM}{%
         family={Head-Gordon},
         familyi={H\bibinitperiod-G\bibinitperiod},
         given={Martin},
         giveni={M\bibinitperiod},
      }}%
    }
    \list{publisher}{1}{%
      {AIP Publishing}%
    }
    \strng{namehash}{SRP+1}
    \strng{fullhash}{SRPDJRASYKJHGM1}
    \field{labelnamesource}{author}
    \field{labeltitlesource}{title}
    \verb{doi}
    \verb 10.1063/1.2234371
    \endverb
    \field{number}{7}
    \field{pages}{074108}
    \field{title}{Dual-basis second-order M{\o}ller-Plesset perturbation
  theory: A reduced-cost reference for correlation calculations}
    \field{volume}{125}
    \field{journaltitle}{J. Chem. Phys.}
    \field{year}{2006}
  \endentry

  \entry{weigend1997ri}{article}{}
    \name{author}{2}{}{%
      {{hash=WF}{%
         family={Weigend},
         familyi={W\bibinitperiod},
         given={Florian},
         giveni={F\bibinitperiod},
      }}%
      {{hash=HM}{%
         family={H{\"a}ser},
         familyi={H\bibinitperiod},
         given={Marco},
         giveni={M\bibinitperiod},
      }}%
    }
    \list{publisher}{1}{%
      {Springer}%
    }
    \strng{namehash}{WFHM1}
    \strng{fullhash}{WFHM1}
    \field{labelnamesource}{author}
    \field{labeltitlesource}{title}
    \verb{doi}
    \verb 10.1007/s002140050269
    \endverb
    \field{number}{1-4}
    \field{pages}{331\bibrangedash 340}
    \field{title}{RI-MP2: first derivatives and global consistency}
    \field{volume}{97}
    \field{journaltitle}{Theor. Chem. Acc.}
    \field{year}{1997}
  \endentry

  \entry{DLPNO-MP2-F12}{article}{}
    \name{author}{5}{}{%
      {{hash=PF}{%
         family={Pavošević},
         familyi={P\bibinitperiod},
         given={Fabijan},
         giveni={F\bibinitperiod},
      }}%
      {{hash=PP}{%
         family={Pinski},
         familyi={P\bibinitperiod},
         given={Peter},
         giveni={P\bibinitperiod},
      }}%
      {{hash=RC}{%
         family={Riplinger},
         familyi={R\bibinitperiod},
         given={Christoph},
         giveni={C\bibinitperiod},
      }}%
      {{hash=NF}{%
         family={Neese},
         familyi={N\bibinitperiod},
         given={Frank},
         giveni={F\bibinitperiod},
      }}%
      {{hash=VEF}{%
         family={Valeev},
         familyi={V\bibinitperiod},
         given={Edward\bibnamedelima F.},
         giveni={E\bibinitperiod\bibinitdelim F\bibinitperiod},
      }}%
    }
    \strng{namehash}{PF+1}
    \strng{fullhash}{PFPPRCNFVEF1}
    \field{labelnamesource}{author}
    \field{labeltitlesource}{title}
    \verb{doi}
    \verb 10.1063/1.4945444
    \endverb
    \verb{eprint}
    \verb https://doi.org/10.1063/1.4945444
    \endverb
    \field{number}{14}
    \field{pages}{144109}
    \field{title}{SparseMaps—A systematic infrastructure for reduced-scaling
  electronic structure methods. IV. Linear-scaling second-order explicitly
  correlated energy with pair natural orbitals}
    \verb{url}
    \verb https://doi.org/10.1063/1.4945444
    \endverb
    \field{volume}{144}
    \field{journaltitle}{J. Chem. Phys.}
    \field{year}{2016}
  \endentry

  \entry{sosdistance}{article}{}
    \name{author}{3}{}{%
      {{hash=LRC}{%
         family={Lochan},
         familyi={L\bibinitperiod},
         given={Rohini\bibnamedelima C.},
         giveni={R\bibinitperiod\bibinitdelim C\bibinitperiod},
      }}%
      {{hash=JY}{%
         family={Jung},
         familyi={J\bibinitperiod},
         given={Yousung},
         giveni={Y\bibinitperiod},
      }}%
      {{hash=HGM}{%
         family={Head-Gordon},
         familyi={H\bibinitperiod-G\bibinitperiod},
         given={Martin},
         giveni={M\bibinitperiod},
      }}%
    }
    \strng{namehash}{LRC+1}
    \strng{fullhash}{LRCJYHGM1}
    \field{labelnamesource}{author}
    \field{labeltitlesource}{title}
    \verb{doi}
    \verb 10.1021/jp0514426
    \endverb
    \verb{eprint}
    \verb https://doi.org/10.1021/jp0514426
    \endverb
    \field{note}{PMID: 16834130}
    \field{number}{33}
    \field{pages}{7598\bibrangedash 7605}
    \field{title}{Scaled Opposite Spin Second Order Møller−Plesset Theory
  with Improved Physical Description of Long-Range Dispersion Interactions}
    \verb{url}
    \verb https://doi.org/10.1021/jp0514426
    \endverb
    \field{volume}{109}
    \field{journaltitle}{J. Phys. Chem. A}
    \field{year}{2005}
  \endentry

  \entry{o2regularisation}{article}{}
    \name{author}{2}{}{%
      {{hash=SD}{%
         family={Stück},
         familyi={S\bibinitperiod},
         given={David},
         giveni={D\bibinitperiod},
      }}%
      {{hash=HGM}{%
         family={Head-Gordon},
         familyi={H\bibinitperiod-G\bibinitperiod},
         given={Martin},
         giveni={M\bibinitperiod},
      }}%
    }
    \strng{namehash}{SDHGM1}
    \strng{fullhash}{SDHGM1}
    \field{labelnamesource}{author}
    \field{labeltitlesource}{title}
    \verb{doi}
    \verb 10.1063/1.4851816
    \endverb
    \verb{eprint}
    \verb https://doi.org/10.1063/1.4851816
    \endverb
    \field{number}{24}
    \field{pages}{244109}
    \field{title}{Regularized orbital-optimized second-order perturbation
  theory}
    \verb{url}
    \verb https://doi.org/10.1063/1.4851816
    \endverb
    \field{volume}{139}
    \field{journaltitle}{J. Chem. Phys.}
    \field{year}{2013}
  \endentry

  \entry{o2regularisationfix}{article}{}
    \name{author}{3}{}{%
      {{hash=RRM}{%
         family={Razban},
         familyi={R\bibinitperiod},
         given={Rostam\bibnamedelima M.},
         giveni={R\bibinitperiod\bibinitdelim M\bibinitperiod},
      }}%
      {{hash=SD}{%
         family={Stück},
         familyi={S\bibinitperiod},
         given={David},
         giveni={D\bibinitperiod},
      }}%
      {{hash=HGM}{%
         family={Head-Gordon},
         familyi={H\bibinitperiod-G\bibinitperiod},
         given={Martin},
         giveni={M\bibinitperiod},
      }}%
    }
    \list{publisher}{1}{%
      {Taylor & Francis}%
    }
    \strng{namehash}{RRM+1}
    \strng{fullhash}{RRMSDHGM1}
    \field{labelnamesource}{author}
    \field{labeltitlesource}{title}
    \verb{doi}
    \verb 10.1080/00268976.2017.1284355
    \endverb
    \verb{eprint}
    \verb https://doi.org/10.1080/00268976.2017.1284355
    \endverb
    \field{number}{17-18}
    \field{pages}{2102\bibrangedash 2109}
    \field{title}{Addressing first derivative discontinuities in
  orbital-optimised opposite-spin scaled second-order perturbation theory with
  regularisation}
    \verb{url}
    \verb https://doi.org/10.1080/00268976.2017.1284355
    \endverb
    \field{volume}{115}
    \field{journaltitle}{Mol. Phys.}
    \field{year}{2017}
  \endentry

  \entry{regoomp22018}{article}{}
    \name{author}{2}{}{%
      {{hash=LJ}{%
         family={Lee},
         familyi={L\bibinitperiod},
         given={Joonho},
         giveni={J\bibinitperiod},
      }}%
      {{hash=HGM}{%
         family={Head-Gordon},
         familyi={H\bibinitperiod-G\bibinitperiod},
         given={Martin},
         giveni={M\bibinitperiod},
      }}%
    }
    \strng{namehash}{LJHGM1}
    \strng{fullhash}{LJHGM1}
    \field{labelnamesource}{author}
    \field{labeltitlesource}{title}
    \verb{doi}
    \verb 10.1021/acs.jctc.8b00731
    \endverb
    \verb{eprint}
    \verb https://doi.org/10.1021/acs.jctc.8b00731
    \endverb
    \field{note}{PMID: 30130398}
    \field{number}{10}
    \field{pages}{5203\bibrangedash 5219}
    \field{title}{Regularized Orbital-Optimized Second-Order Møller–Plesset
  Perturbation Theory: A Reliable Fifth-Order-Scaling Electron Correlation
  Model with Orbital Energy Dependent Regularizers}
    \verb{url}
    \verb https://doi.org/10.1021/acs.jctc.8b00731
    \endverb
    \field{volume}{14}
    \field{journaltitle}{J. Chem. Theory Comput.}
    \field{year}{2018}
  \endentry

  \entry{bsseatenuation}{article}{}
    \name{author}{2}{}{%
      {{hash=GM}{%
         family={Goldey},
         familyi={G\bibinitperiod},
         given={Matthew},
         giveni={M\bibinitperiod},
      }}%
      {{hash=HGM}{%
         family={Head-Gordon},
         familyi={H\bibinitperiod-G\bibinitperiod},
         given={Martin},
         giveni={M\bibinitperiod},
      }}%
    }
    \strng{namehash}{GMHGM1}
    \strng{fullhash}{GMHGM1}
    \field{labelnamesource}{author}
    \field{labeltitlesource}{title}
    \verb{doi}
    \verb 10.1021/jz301694b
    \endverb
    \verb{eprint}
    \verb https://doi.org/10.1021/jz301694b
    \endverb
    \field{note}{PMID: 26290993}
    \field{number}{23}
    \field{pages}{3592\bibrangedash 3598}
    \field{title}{Attenuating Away the Errors in Inter- and Intramolecular
  Interactions from Second-Order Møller–Plesset Calculations in the Small
  Aug-cc-pVDZ Basis Set}
    \verb{url}
    \verb https://doi.org/10.1021/jz301694b
    \endverb
    \field{volume}{3}
    \field{journaltitle}{J. Phys. Chem. Lett.}
    \field{year}{2012}
  \endentry

  \entry{attenuatedTZ}{article}{}
    \name{author}{3}{}{%
      {{hash=GM}{%
         family={Goldey},
         familyi={G\bibinitperiod},
         given={Matthew},
         giveni={M\bibinitperiod},
      }}%
      {{hash=DA}{%
         family={Dutoi},
         familyi={D\bibinitperiod},
         given={Anthony},
         giveni={A\bibinitperiod},
      }}%
      {{hash=HGM}{%
         family={Head-Gordon},
         familyi={H\bibinitperiod-G\bibinitperiod},
         given={Martin},
         giveni={M\bibinitperiod},
      }}%
    }
    \list{publisher}{1}{%
      {The Royal Society of Chemistry}%
    }
    \strng{namehash}{GM+1}
    \strng{fullhash}{GMDAHGM1}
    \field{labelnamesource}{author}
    \field{labeltitlesource}{title}
    \verb{doi}
    \verb 10.1039/C3CP51826D
    \endverb
    \field{issue}{38}
    \field{pages}{15869\bibrangedash 15875}
    \field{title}{Attenuated second-order Møller–Plesset perturbation
  theory: performance within the aug-cc-pVTZ basis}
    \field{volume}{15}
    \field{journaltitle}{Phys. Chem. Chem. Phys.}
    \field{year}{2013}
  \endentry

  \entry{oo-mp2}{article}{}
    \name{author}{5}{}{%
      {{hash=BU}{%
         family={Bozkaya},
         familyi={B\bibinitperiod},
         given={Uğur},
         giveni={U\bibinitperiod},
      }}%
      {{hash=TJM}{%
         family={Turney},
         familyi={T\bibinitperiod},
         given={Justin\bibnamedelima M.},
         giveni={J\bibinitperiod\bibinitdelim M\bibinitperiod},
      }}%
      {{hash=YY}{%
         family={Yamaguchi},
         familyi={Y\bibinitperiod},
         given={Yukio},
         giveni={Y\bibinitperiod},
      }}%
      {{hash=SHF}{%
         family={Schaefer},
         familyi={S\bibinitperiod},
         given={Henry\bibnamedelima F.},
         giveni={H\bibinitperiod\bibinitdelim F\bibinitperiod},
      }}%
      {{hash=SCD}{%
         family={Sherrill},
         familyi={S\bibinitperiod},
         given={C.\bibnamedelima David},
         giveni={C\bibinitperiod\bibinitdelim D\bibinitperiod},
      }}%
    }
    \strng{namehash}{BU+1}
    \strng{fullhash}{BUTJMYYSHFSCD1}
    \field{labelnamesource}{author}
    \field{labeltitlesource}{title}
    \verb{doi}
    \verb 10.1063/1.3631129
    \endverb
    \verb{eprint}
    \verb https://doi.org/10.1063/1.3631129
    \endverb
    \field{number}{10}
    \field{pages}{104103}
    \field{title}{Quadratically convergent algorithm for orbital optimization
  in the orbital-optimized coupled-cluster doubles method and in
  orbital-optimized second-order Møller-Plesset perturbation theory}
    \verb{url}
    \verb https://doi.org/10.1063/1.3631129
    \endverb
    \field{volume}{135}
    \field{journaltitle}{J. Chem. Phys.}
    \field{year}{2011}
  \endentry

  \entry{oo-sos-mp2}{article}{}
    \name{author}{2}{}{%
      {{hash=LRC}{%
         family={Lochan},
         familyi={L\bibinitperiod},
         given={Rohini\bibnamedelima C.},
         giveni={R\bibinitperiod\bibinitdelim C\bibinitperiod},
      }}%
      {{hash=HGM}{%
         family={Head-Gordon},
         familyi={H\bibinitperiod-G\bibinitperiod},
         given={Martin},
         giveni={M\bibinitperiod},
      }}%
    }
    \strng{namehash}{LRCHGM1}
    \strng{fullhash}{LRCHGM1}
    \field{labelnamesource}{author}
    \field{labeltitlesource}{title}
    \verb{doi}
    \verb 10.1063/1.2718952
    \endverb
    \verb{eprint}
    \verb https://doi.org/10.1063/1.2718952
    \endverb
    \field{number}{16}
    \field{pages}{164101}
    \field{title}{Orbital-optimized opposite-spin scaled second-order
  correlation: An economical method to improve the description of open-shell
  molecules}
    \verb{url}
    \verb https://doi.org/10.1063/1.2718952
    \endverb
    \field{volume}{126}
    \field{journaltitle}{J. Chem. Phys.}
    \field{year}{2007}
  \endentry

  \entry{SCS-MP2mod}{article}{}
    \name{author}{6}{}{%
      {{hash=CI}{%
         family={Cacelli},
         familyi={C\bibinitperiod},
         given={Ivo},
         giveni={I\bibinitperiod},
      }}%
      {{hash=LF}{%
         family={Lipparini},
         familyi={L\bibinitperiod},
         given={Filippo},
         giveni={F\bibinitperiod},
      }}%
      {{hash=GdSL}{%
         prefix={Greff\bibnamedelima da},
         prefixi={G\bibinitperiod\bibinitdelim d\bibinitperiod},
         family={Silveira},
         familyi={S\bibinitperiod},
         given={Leandro},
         giveni={L\bibinitperiod},
      }}%
      {{hash=JM}{%
         family={Jacobs},
         familyi={J\bibinitperiod},
         given={Matheus},
         giveni={M\bibinitperiod},
      }}%
      {{hash=LPR}{%
         family={Livotto},
         familyi={L\bibinitperiod},
         given={Paolo\bibnamedelima Roberto},
         giveni={P\bibinitperiod\bibinitdelim R\bibinitperiod},
      }}%
      {{hash=PG}{%
         family={Prampolini},
         familyi={P\bibinitperiod},
         given={Giacomo},
         giveni={G\bibinitperiod},
      }}%
    }
    \strng{namehash}{CI+1}
    \strng{fullhash}{CILFSLGdJMLPRPG1}
    \field{labelnamesource}{author}
    \field{labeltitlesource}{title}
    \verb{doi}
    \verb 10.1063/1.5094288
    \endverb
    \verb{eprint}
    \verb https://doi.org/10.1063/1.5094288
    \endverb
    \field{number}{23}
    \field{pages}{234113}
    \field{title}{Accurate interaction energies by spin component scaled
  Möller-Plesset second order perturbation theory calculations with optimized
  basis sets (SCS-MP2mod): Development and application to aromatic
  heterocycles}
    \verb{url}
    \verb https://doi.org/10.1063/1.5094288
    \endverb
    \field{volume}{150}
    \field{journaltitle}{J. Chem. Phys.}
    \field{year}{2019}
  \endentry

  \entry{petersson1985complete}{article}{}
    \name{author}{3}{}{%
      {{hash=PG}{%
         family={Petersson},
         familyi={P\bibinitperiod},
         given={GA},
         giveni={G\bibinitperiod},
      }}%
      {{hash=YAK}{%
         family={Yee},
         familyi={Y\bibinitperiod},
         given={Arnold\bibnamedelima K},
         giveni={A\bibinitperiod\bibinitdelim K\bibinitperiod},
      }}%
      {{hash=BA}{%
         family={Bennett},
         familyi={B\bibinitperiod},
         given={Andrew},
         giveni={A\bibinitperiod},
      }}%
    }
    \list{publisher}{1}{%
      {AIP Publishing}%
    }
    \strng{namehash}{PG+1}
    \strng{fullhash}{PGYAKBA1}
    \field{labelnamesource}{author}
    \field{labeltitlesource}{title}
    \verb{doi}
    \verb 10.1063/1.449724
    \endverb
    \field{number}{10}
    \field{pages}{5105\bibrangedash 5128}
    \field{title}{Complete basis set correlation energies. III. The total
  correlation energy of the neon atom}
    \field{volume}{83}
    \field{journaltitle}{J. Chem. Phys.}
    \field{year}{1985}
  \endentry

  \entry{szabo1996modern}{book}{}
    \name{author}{1}{}{%
      {{hash=SA}{%
         family={Szabo},
         familyi={S\bibinitperiod},
         given={Attila},
         giveni={A\bibinitperiod},
      }}%
    }
    \list{publisher}{1}{%
      {Dover Publications}%
    }
    \strng{namehash}{SA2}
    \strng{fullhash}{SA2}
    \field{labelnamesource}{author}
    \field{labeltitlesource}{title}
  \field{isbn}{\href{https://isbnsearch.org/isbn/9780486691862}{978-0486691862}}
    \field{title}{Modern quantum chemistry: introduction to advanced electronic
  structure theory}
    \list{location}{1}{%
      {Mineola, N.Y}%
    }
    \field{year}{1996}
  \endentry

  \entry{pople1987quadratic}{article}{}
    \name{author}{3}{}{%
      {{hash=PJA}{%
         family={Pople},
         familyi={P\bibinitperiod},
         given={John\bibnamedelima A},
         giveni={J\bibinitperiod\bibinitdelim A\bibinitperiod},
      }}%
      {{hash=HGM}{%
         family={Head-Gordon},
         familyi={H\bibinitperiod-G\bibinitperiod},
         given={Martin},
         giveni={M\bibinitperiod},
      }}%
      {{hash=RK}{%
         family={Raghavachari},
         familyi={R\bibinitperiod},
         given={Krishnan},
         giveni={K\bibinitperiod},
      }}%
    }
    \list{publisher}{1}{%
      {AIP Publishing}%
    }
    \strng{namehash}{PJA+1}
    \strng{fullhash}{PJAHGMRK1}
    \field{labelnamesource}{author}
    \field{labeltitlesource}{title}
    \verb{doi}
    \verb 10.1063/1.453520
    \endverb
    \field{number}{10}
    \field{pages}{5968\bibrangedash 5975}
    \field{title}{Quadratic configuration interaction. A general technique for
  determining electron correlation energies}
    \field{volume}{87}
    \field{journaltitle}{J. Chem. Phys.}
    \field{year}{1987}
  \endentry

  \entry{grimme2003improved}{article}{}
    \name{author}{1}{}{%
      {{hash=GS}{%
         family={Grimme},
         familyi={G\bibinitperiod},
         given={Stefan},
         giveni={S\bibinitperiod},
      }}%
    }
    \list{publisher}{1}{%
      {AIP Publishing}%
    }
    \strng{namehash}{GS1}
    \strng{fullhash}{GS1}
    \field{labelnamesource}{author}
    \field{labeltitlesource}{title}
    \verb{doi}
    \verb 10.1063/1.1569242
    \endverb
    \field{number}{20}
    \field{pages}{9095\bibrangedash 9102}
    \field{title}{Improved second-order M{\o}ller--Plesset perturbation theory
  by separate scaling of parallel-and antiparallel-spin pair correlation
  energies}
    \field{volume}{118}
    \field{journaltitle}{J. Chem. Phys.}
    \field{year}{2003}
  \endentry

  \entry{Szabados2006}{article}{}
    \name{author}{1}{}{%
      {{hash=SA}{%
         family={Szabados},
         familyi={S\bibinitperiod},
         given={Agnes},
         giveni={A\bibinitperiod},
      }}%
    }
    \strng{namehash}{SA1}
    \strng{fullhash}{SA1}
    \field{labelnamesource}{author}
    \field{labeltitlesource}{title}
    \verb{doi}
    \verb 10.1063/1.2404660
    \endverb
    \verb{eprint}
    \verb https://doi.org/10.1063/1.2404660
    \endverb
    \field{number}{21}
    \field{pages}{214105}
    \field{title}{Theoretical interpretation of Grimme's spin-component-scaled
  second order Møller-Plesset theory}
    \verb{url}
    \verb https://doi.org/10.1063/1.2404660
    \endverb
    \field{volume}{125}
    \field{journaltitle}{J. Chem. Phys.}
    \field{year}{2006}
  \endentry

  \entry{fink2010}{article}{}
    \name{author}{1}{}{%
      {{hash=FRF}{%
         family={Fink},
         familyi={F\bibinitperiod},
         given={Reinhold\bibnamedelima F.},
         giveni={R\bibinitperiod\bibinitdelim F\bibinitperiod},
      }}%
    }
    \strng{namehash}{FRF1}
    \strng{fullhash}{FRF1}
    \field{labelnamesource}{author}
    \field{labeltitlesource}{title}
    \verb{doi}
    \verb 10.1063/1.3503041
    \endverb
    \verb{eprint}
    \verb https://doi.org/10.1063/1.3503041
    \endverb
    \field{number}{17}
    \field{pages}{174113}
    \field{title}{Spin-component-scaled Møller–Plesset (SCS-MP) perturbation
  theory: A generalization of the MP approach with improved properties}
    \verb{url}
    \verb https://doi.org/10.1063/1.3503041
    \endverb
    \field{volume}{133}
    \field{journaltitle}{J. Chem. Phys.}
    \field{year}{2010}
  \endentry

  \entry{jung2004scaled}{article}{}
    \name{author}{4}{}{%
      {{hash=JY}{%
         family={Jung},
         familyi={J\bibinitperiod},
         given={Yousung},
         giveni={Y\bibinitperiod},
      }}%
      {{hash=LRC}{%
         family={Lochan},
         familyi={L\bibinitperiod},
         given={Rohini\bibnamedelima C},
         giveni={R\bibinitperiod\bibinitdelim C\bibinitperiod},
      }}%
      {{hash=DAD}{%
         family={Dutoi},
         familyi={D\bibinitperiod},
         given={Anthony\bibnamedelima D},
         giveni={A\bibinitperiod\bibinitdelim D\bibinitperiod},
      }}%
      {{hash=HGM}{%
         family={Head-Gordon},
         familyi={H\bibinitperiod-G\bibinitperiod},
         given={Martin},
         giveni={M\bibinitperiod},
      }}%
    }
    \list{publisher}{1}{%
      {AIP Publishing}%
    }
    \strng{namehash}{JY+1}
    \strng{fullhash}{JYLRCDADHGM1}
    \field{labelnamesource}{author}
    \field{labeltitlesource}{title}
    \verb{doi}
    \verb 10.1063/1.1809602
    \endverb
    \field{number}{20}
    \field{pages}{9793\bibrangedash 9802}
    \field{title}{Scaled opposite-spin second order M{\o}ller--Plesset
  correlation energy: an economical electronic structure method}
    \field{volume}{121}
    \field{journaltitle}{J. Chem. Phys.}
    \field{year}{2004}
  \endentry

  \entry{neese2009assessment}{article}{}
    \name{author}{5}{}{%
      {{hash=NF}{%
         family={Neese},
         familyi={N\bibinitperiod},
         given={Frank},
         giveni={F\bibinitperiod},
      }}%
      {{hash=ST}{%
         family={Schwabe},
         familyi={S\bibinitperiod},
         given={Tobias},
         giveni={T\bibinitperiod},
      }}%
      {{hash=KS}{%
         family={Kossmann},
         familyi={K\bibinitperiod},
         given={Simone},
         giveni={S\bibinitperiod},
      }}%
      {{hash=SB}{%
         family={Schirmer},
         familyi={S\bibinitperiod},
         given={Birgitta},
         giveni={B\bibinitperiod},
      }}%
      {{hash=GS}{%
         family={Grimme},
         familyi={G\bibinitperiod},
         given={Stefan},
         giveni={S\bibinitperiod},
      }}%
    }
    \list{publisher}{1}{%
      {ACS Publications}%
    }
    \strng{namehash}{NF+1}
    \strng{fullhash}{NFSTKSSBGS1}
    \field{labelnamesource}{author}
    \field{labeltitlesource}{title}
    \verb{doi}
    \verb 10.1021/ct9003299
    \endverb
    \field{number}{11}
    \field{pages}{3060\bibrangedash 3073}
    \field{title}{Assessment of orbital-optimized, spin-component scaled
  second-order many-body perturbation theory for thermochemistry and kinetics}
    \field{volume}{5}
    \field{journaltitle}{J. Chem. Theory Comput.}
    \field{year}{2009}
  \endentry

  \entry{szabados2011}{article}{}
    \name{author}{2}{}{%
      {{hash=ÃS}{%
         prefix={Ágnes},
         prefixi={Ã\bibinitperiod},
         family={Szabados},
         familyi={S\bibinitperiod},
      }}%
      {{hash=NP}{%
         family={Nagy},
         familyi={N\bibinitperiod},
         given={Péter},
         giveni={P\bibinitperiod},
      }}%
    }
    \strng{namehash}{SÃNP1}
    \strng{fullhash}{SÃNP1}
    \field{labelnamesource}{author}
    \field{labeltitlesource}{title}
    \verb{doi}
    \verb 10.1021/jp108575a
    \endverb
    \verb{eprint}
    \verb http://dx.doi.org/10.1021/jp108575a
    \endverb
    \field{note}{PMID: 21190320}
    \field{number}{4}
    \field{pages}{523\bibrangedash 534}
    \field{title}{Spin Component Scaling in Multiconfiguration Perturbation
  Theory}
    \verb{url}
    \verb http://dx.doi.org/10.1021/jp108575a
    \endverb
    \field{volume}{115}
    \field{journaltitle}{J. Phys. Chem. A}
    \field{year}{2011}
  \endentry

  \entry{distasio2007optimized}{article}{}
    \name{author}{2}{}{%
      {{hash=DJRA}{%
         family={Distasio\bibnamedelima Jr},
         familyi={D\bibinitperiod\bibinitdelim J\bibinitperiod},
         given={Robert\bibnamedelima A},
         giveni={R\bibinitperiod\bibinitdelim A\bibinitperiod},
      }}%
      {{hash=HGM}{%
         family={Head-Gordon},
         familyi={H\bibinitperiod-G\bibinitperiod},
         given={Martin},
         giveni={M\bibinitperiod},
      }}%
    }
    \list{publisher}{1}{%
      {Taylor \& Francis}%
    }
    \strng{namehash}{DJRAHGM1}
    \strng{fullhash}{DJRAHGM1}
    \field{labelnamesource}{author}
    \field{labeltitlesource}{title}
    \verb{doi}
    \verb 10.1080/00268970701283781
    \endverb
    \field{number}{8}
    \field{pages}{1073\bibrangedash 1083}
    \field{title}{Optimized spin-component scaled second-order
  M{\o}ller-Plesset perturbation theory for intermolecular interaction
  energies}
    \field{volume}{105}
    \field{journaltitle}{Mol. Phys.}
    \field{year}{2007}
  \endentry

  \entry{hill2007spin}{article}{}
    \name{author}{2}{}{%
      {{hash=HJG}{%
         family={Hill},
         familyi={H\bibinitperiod},
         given={J\bibnamedelima Grant},
         giveni={J\bibinitperiod\bibinitdelim G\bibinitperiod},
      }}%
      {{hash=PJA}{%
         family={Platts},
         familyi={P\bibinitperiod},
         given={James\bibnamedelima A},
         giveni={J\bibinitperiod\bibinitdelim A\bibinitperiod},
      }}%
    }
    \list{publisher}{1}{%
      {ACS Publications}%
    }
    \strng{namehash}{HJGPJA1}
    \strng{fullhash}{HJGPJA1}
    \field{labelnamesource}{author}
    \field{labeltitlesource}{title}
    \verb{doi}
    \verb 10.1021/ct6002737
    \endverb
    \field{number}{1}
    \field{pages}{80\bibrangedash 85}
    \field{title}{Spin-component scaling methods for weak and stacking
  interactions}
    \field{volume}{3}
    \field{journaltitle}{J. Chem. Theory Comput.}
    \field{year}{2007}
  \endentry

  \entry{SCSILMP2}{article}{}
    \name{author}{2}{}{%
      {{hash=RJ}{%
         family={Rigby},
         familyi={R\bibinitperiod},
         given={Jason},
         giveni={J\bibinitperiod},
      }}%
      {{hash=IEI}{%
         family={Izgorodina},
         familyi={I\bibinitperiod},
         given={Ekaterina\bibnamedelima I.},
         giveni={E\bibinitperiod\bibinitdelim I\bibinitperiod},
      }}%
    }
    \strng{namehash}{RJIEI1}
    \strng{fullhash}{RJIEI1}
    \field{labelnamesource}{author}
    \field{labeltitlesource}{title}
    \verb{doi}
    \verb 10.1021/ct500309x
    \endverb
    \verb{eprint}
    \verb http://dx.doi.org/10.1021/ct500309x
    \endverb
    \field{note}{PMID: 26588282}
    \field{number}{8}
    \field{pages}{3111\bibrangedash 3122}
    \field{title}{New SCS- and SOS-MP2 Coefficients Fitted to Semi-Coulombic
  Systems}
    \verb{url}
    \verb http://dx.doi.org/10.1021/ct500309x
    \endverb
    \field{volume}{10}
    \field{journaltitle}{J. Chem. Theory Comput.}
    \field{year}{2014}
  \endentry

  \entry{Gordon2014}{article}{}
    \name{author}{2}{}{%
      {{hash=GM}{%
         family={Goldey},
         familyi={G\bibinitperiod},
         given={Matthew},
         giveni={M\bibinitperiod},
      }}%
      {{hash=HGM}{%
         family={Head-Gordon},
         familyi={H\bibinitperiod-G\bibinitperiod},
         given={Martin},
         giveni={M\bibinitperiod},
      }}%
    }
    \strng{namehash}{GMHGM1}
    \strng{fullhash}{GMHGM1}
    \field{labelnamesource}{author}
    \field{labeltitlesource}{title}
    \verb{doi}
    \verb 10.1021/jp4126478
    \endverb
    \field{number}{24}
    \field{pages}{6519\bibrangedash 6525}
    \field{title}{Separate Electronic Attenuation Allowing a
  Spin-Component-Scaled Second-Order Møller–Plesset Theory to Be Effective
  for Both Thermochemistry and Noncovalent Interactions}
    \field{volume}{118}
    \field{journaltitle}{J. Phys. Chem. B}
    \field{year}{2014}
  \endentry

  \entry{grimme_review_2012}{article}{}
    \name{author}{3}{}{%
      {{hash=GS}{%
         family={Grimme},
         familyi={G\bibinitperiod},
         given={Stefan},
         giveni={S\bibinitperiod},
      }}%
      {{hash=GL}{%
         family={Goerigk},
         familyi={G\bibinitperiod},
         given={Lars},
         giveni={L\bibinitperiod},
      }}%
      {{hash=FRF}{%
         family={Fink},
         familyi={F\bibinitperiod},
         given={Reinhold\bibnamedelima F.},
         giveni={R\bibinitperiod\bibinitdelim F\bibinitperiod},
      }}%
    }
    \strng{namehash}{GS+1}
    \strng{fullhash}{GSGLFRF1}
    \field{labelnamesource}{author}
    \field{labeltitlesource}{title}
    \verb{doi}
    \verb 10.1002/wcms.1110
    \endverb
    \verb{eprint}
    \verb https://onlinelibrary.wiley.com/doi/pdf/10.1002/wcms.1110
    \endverb
    \field{number}{6}
    \field{pages}{886\bibrangedash 906}
    \field{title}{Spin-component-scaled electron correlation methods}
    \verb{url}
    \verb https://onlinelibrary.wiley.com/doi/abs/10.1002/wcms.1110
    \endverb
    \field{volume}{2}
    \field{journaltitle}{WIREs Comput. Mol. Sci.}
    \field{year}{2012}
  \endentry

  \entry{martin2019}{article}{}
    \name{author}{2}{}{%
      {{hash=SN}{%
         family={Sylvetsky},
         familyi={S\bibinitperiod},
         given={Nitai},
         giveni={N\bibinitperiod},
      }}%
      {{hash=MJML}{%
         family={Martin},
         familyi={M\bibinitperiod},
         given={Jan M.\bibnamedelima L.},
         giveni={J\bibinitperiod\bibinitdelim M\bibinitperiod\bibinitdelim
  L\bibinitperiod},
      }}%
    }
    \list{publisher}{1}{%
      {Taylor & Francis}%
    }
    \strng{namehash}{SNMJML1}
    \strng{fullhash}{SNMJML1}
    \field{labelnamesource}{author}
    \field{labeltitlesource}{title}
    \verb{doi}
    \verb 10.1080/00268976.2018.1478140
    \endverb
    \verb{eprint}
    \verb https://doi.org/10.1080/00268976.2018.1478140
    \endverb
    \field{number}{9-12}
    \field{pages}{1078\bibrangedash 1087}
    \field{title}{Probing the basis set limit for thermochemical contributions
  of inner-shell correlation: balance of core-core and core-valence
  contributions}
    \verb{url}
    \verb https://doi.org/10.1080/00268976.2018.1478140
    \endverb
    \field{volume}{117}
    \field{journaltitle}{Mol. Phys.}
    \field{year}{2019}
  \endentry

  \entry{dlpno-ccsdt-core}{article}{}
    \name{author}{6}{}{%
      {{hash=MY}{%
         family={Minenkov},
         familyi={M\bibinitperiod},
         given={Yury},
         giveni={Y\bibinitperiod},
      }}%
      {{hash=BG}{%
         family={Bistoni},
         familyi={B\bibinitperiod},
         given={Giovanni},
         giveni={G\bibinitperiod},
      }}%
      {{hash=RC}{%
         family={Riplinger},
         familyi={R\bibinitperiod},
         given={Christoph},
         giveni={C\bibinitperiod},
      }}%
      {{hash=AAA}{%
         family={Auer},
         familyi={A\bibinitperiod},
         given={Alexander\bibnamedelima A.},
         giveni={A\bibinitperiod\bibinitdelim A\bibinitperiod},
      }}%
      {{hash=NF}{%
         family={Neese},
         familyi={N\bibinitperiod},
         given={Frank},
         giveni={F\bibinitperiod},
      }}%
      {{hash=CL}{%
         family={Cavallo},
         familyi={C\bibinitperiod},
         given={Luigi},
         giveni={L\bibinitperiod},
      }}%
    }
    \list{publisher}{1}{%
      {The Royal Society of Chemistry}%
    }
    \strng{namehash}{MY+1}
    \strng{fullhash}{MYBGRCAAANFCL1}
    \field{labelnamesource}{author}
    \field{labeltitlesource}{title}
    \verb{doi}
    \verb 10.1039/C7CP00836H
    \endverb
    \field{issue}{14}
    \field{pages}{9374\bibrangedash 9391}
    \field{title}{Pair natural orbital and canonical coupled cluster reaction
  enthalpies involving light to heavy alkali and alkaline earth metals: the
  importance of sub-valence correlation}
    \verb{url}
    \verb http://dx.doi.org/10.1039/C7CP00836H
    \endverb
    \field{volume}{19}
    \field{journaltitle}{Phys. Chem. Chem. Phys.}
    \field{year}{2017}
  \endentry

  \entry{AHLRICHS197931}{article}{}
    \name{author}{1}{}{%
      {{hash=AR}{%
         family={Ahlrichs},
         familyi={A\bibinitperiod},
         given={Reinhart},
         giveni={R\bibinitperiod},
      }}%
    }
    \strng{namehash}{AR1}
    \strng{fullhash}{AR1}
    \field{labelnamesource}{author}
    \field{labeltitlesource}{title}
    \field{abstract}{%
    The CEPA equations are derived and the connections between CEPA and CI or
  MBPT procedures are discussed in detail. A new self-consistent direct method
  (SCD) is proposed to treat equations which include: complete CI(SD), CEPA,
  MP-PT (3rd order), and MBPT of Bartlett and Silver as special cases.
  Applications prove the excellent quality of the CEPA versions CEPA(1) and
  CEPA(2).%
    }
    \verb{doi}
    \verb https://doi.org/10.1016/0010-4655(79)90067-5
    \endverb
    \field{issn}{0010-4655}
    \field{number}{1}
    \field{pages}{31 \bibrangedash  45}
    \field{title}{Many body perturbation calculations and coupled electron pair
  models}
    \verb{url}
    \verb http://www.sciencedirect.com/science/article/pii/0010465579900675
    \endverb
    \field{volume}{17}
    \field{journaltitle}{Comput. Phys. Commun.}
    \field{year}{1979}
  \endentry

  \entry{WENNMOHS2008217}{article}{}
    \name{author}{2}{}{%
      {{hash=WF}{%
         family={Wennmohs},
         familyi={W\bibinitperiod},
         given={Frank},
         giveni={F\bibinitperiod},
      }}%
      {{hash=NF}{%
         family={Neese},
         familyi={N\bibinitperiod},
         given={Frank},
         giveni={F\bibinitperiod},
      }}%
    }
    \strng{namehash}{WFNF1}
    \strng{fullhash}{WFNF1}
    \field{labelnamesource}{author}
    \field{labeltitlesource}{title}
    \verb{doi}
    \verb https://doi.org/10.1016/j.chemphys.2007.07.001
    \endverb
    \field{issn}{0301-0104}
    \field{note}{Theoretical Spectroscopy and its Impact on Experiment}
    \field{number}{2}
    \field{pages}{217 \bibrangedash  230}
    \field{title}{A comparative study of single reference correlation methods
  of the coupled-pair type}
    \verb{url}
    \verb http://www.sciencedirect.com/science/article/pii/S0301010407002698
    \endverb
    \field{volume}{343}
    \field{journaltitle}{Chem. Phys.}
    \field{year}{2008}
  \endentry

  \entry{RevModPhys.32.179}{article}{}
    \name{author}{1}{}{%
      {{hash=RCCJ}{%
         family={Roothaan},
         familyi={R\bibinitperiod},
         given={C.\bibnamedelima C.\bibnamedelima J.},
         giveni={C\bibinitperiod\bibinitdelim C\bibinitperiod\bibinitdelim
  J\bibinitperiod},
      }}%
    }
    \list{publisher}{1}{%
      {American Physical Society}%
    }
    \strng{namehash}{RCCJ1}
    \strng{fullhash}{RCCJ1}
    \field{labelnamesource}{author}
    \field{labeltitlesource}{title}
    \verb{doi}
    \verb 10.1103/RevModPhys.32.179
    \endverb
    \field{issue}{2}
    \field{pages}{179\bibrangedash 185}
    \field{title}{Self-Consistent Field Theory for Open Shells of Electronic
  Systems}
    \verb{url}
    \verb http://link.aps.org/doi/10.1103/RevModPhys.32.179
    \endverb
    \field{volume}{32}
    \field{journaltitle}{Rev. Mod. Phys.}
    \field{year}{1960}
    \warn{\item Invalid format of field 'month'}
  \endentry

  \entry{Knowles_ccsd_t}{article}{}
    \name{author}{3}{}{%
      {{hash=KPJ}{%
         family={Knowles},
         familyi={K\bibinitperiod},
         given={Peter\bibnamedelima J.},
         giveni={P\bibinitperiod\bibinitdelim J\bibinitperiod},
      }}%
      {{hash=HC}{%
         family={Hampel},
         familyi={H\bibinitperiod},
         given={Claudia},
         giveni={C\bibinitperiod},
      }}%
      {{hash=WH}{%
         family={Werner},
         familyi={W\bibinitperiod},
         given={Hans‐Joachim},
         giveni={H\bibinitperiod},
      }}%
    }
    \strng{namehash}{KPJ+1}
    \strng{fullhash}{KPJHCWH1}
    \field{labelnamesource}{author}
    \field{labeltitlesource}{title}
    \verb{doi}
    \verb 10.1063/1.465990
    \endverb
    \verb{eprint}
    \verb https://doi.org/10.1063/1.465990
    \endverb
    \field{number}{7}
    \field{pages}{5219\bibrangedash 5227}
    \field{title}{Coupled cluster theory for high spin, open shell reference
  wave functions}
    \verb{url}
    \verb https://doi.org/10.1063/1.465990
    \endverb
    \field{volume}{99}
    \field{journaltitle}{J. Chem. Phys}
    \field{year}{1993}
  \endentry

  \entry{errata_knowles}{article}{}
    \name{author}{3}{}{%
      {{hash=KPJ}{%
         family={Knowles},
         familyi={K\bibinitperiod},
         given={Peter\bibnamedelima J.},
         giveni={P\bibinitperiod\bibinitdelim J\bibinitperiod},
      }}%
      {{hash=HC}{%
         family={Hampel},
         familyi={H\bibinitperiod},
         given={Claudia},
         giveni={C\bibinitperiod},
      }}%
      {{hash=WHJ}{%
         family={Werner},
         familyi={W\bibinitperiod},
         given={Hans-Joachim},
         giveni={H\bibinitperiod-J\bibinitperiod},
      }}%
    }
    \strng{namehash}{KPJ+1}
    \strng{fullhash}{KPJHCWHJ1}
    \field{labelnamesource}{author}
    \field{labeltitlesource}{title}
    \verb{doi}
    \verb 10.1063/1.480886
    \endverb
    \verb{eprint}
    \verb https://doi.org/10.1063/1.480886
    \endverb
    \field{number}{6}
    \field{pages}{3106\bibrangedash 3107}
    \field{title}{Erratum: "Coupled cluster theory for high spin, open shell
  reference wave functions" [ J. Chem. Phys. 99, 5219 (1993)]}
    \verb{url}
    \verb https://doi.org/10.1063/1.480886
    \endverb
    \field{volume}{112}
    \field{journaltitle}{J. Chem. Phys}
    \field{year}{2000}
  \endentry

  \entry{peterson2002accurate}{article}{}
    \name{author}{2}{}{%
      {{hash=PKA}{%
         family={Peterson},
         familyi={P\bibinitperiod},
         given={Kirk\bibnamedelima A.},
         giveni={K\bibinitperiod\bibinitdelim A\bibinitperiod},
      }}%
      {{hash=DTH}{%
         family={Dunning},
         familyi={D\bibinitperiod},
         given={Thom\bibnamedelima H.},
         giveni={T\bibinitperiod\bibinitdelim H\bibinitperiod},
      }}%
    }
    \strng{namehash}{PKADTH1}
    \strng{fullhash}{PKADTH1}
    \field{labelnamesource}{author}
    \field{labeltitlesource}{title}
    \verb{doi}
    \verb 10.1063/1.1520138
    \endverb
    \verb{eprint}
    \verb https://doi.org/10.1063/1.1520138
    \endverb
    \field{number}{23}
    \field{pages}{10548\bibrangedash 10560}
    \field{title}{Accurate correlation consistent basis sets for molecular
  core–valence correlation effects: The second row atoms Al–Ar, and the
  first row atoms B–Ne revisited}
    \verb{url}
    \verb https://doi.org/10.1063/1.1520138
    \endverb
    \field{volume}{117}
    \field{journaltitle}{J. Chem. Phys.}
    \field{year}{2002}
  \endentry

  \entry{dunning1989gaussian}{article}{}
    \name{author}{1}{}{%
      {{hash=DTH}{%
         family={Dunning},
         familyi={D\bibinitperiod},
         given={Thom\bibnamedelima H.},
         giveni={T\bibinitperiod\bibinitdelim H\bibinitperiod},
      }}%
    }
    \strng{namehash}{DTH1}
    \strng{fullhash}{DTH1}
    \field{labelnamesource}{author}
    \field{labeltitlesource}{title}
    \verb{doi}
    \verb 10.1063/1.456153
    \endverb
    \verb{eprint}
    \verb https://doi.org/10.1063/1.456153
    \endverb
    \field{number}{2}
    \field{pages}{1007\bibrangedash 1023}
    \field{title}{Gaussian basis sets for use in correlated molecular
  calculations. I. The atoms boron through neon and hydrogen}
    \verb{url}
    \verb https://doi.org/10.1063/1.456153
    \endverb
    \field{volume}{90}
    \field{journaltitle}{J. Chem. Phys.}
    \field{year}{1989}
  \endentry

  \entry{iron2003alkali}{article}{}
    \name{author}{3}{}{%
      {{hash=IMA}{%
         family={Iron},
         familyi={I\bibinitperiod},
         given={Mark\bibnamedelima A},
         giveni={M\bibinitperiod\bibinitdelim A\bibinitperiod},
      }}%
      {{hash=OM}{%
         family={Oren},
         familyi={O\bibinitperiod},
         given={Mikhal},
         giveni={M\bibinitperiod},
      }}%
      {{hash=MJM}{%
         family={Martin},
         familyi={M\bibinitperiod},
         given={Jan\bibnamedelima ML},
         giveni={J\bibinitperiod\bibinitdelim M\bibinitperiod},
      }}%
    }
    \strng{namehash}{IMA+1}
    \strng{fullhash}{IMAOMMJM1}
    \field{labelnamesource}{author}
    \field{labeltitlesource}{title}
    \verb{doi}
    \verb 10.1080/0026897031000094498
    \endverb
    \field{number}{9}
    \field{pages}{1345\bibrangedash 1361}
    \field{title}{Alkali and alkaline earth metal compounds: core—valence
  basis sets and importance of subvalence correlation}
    \field{volume}{101}
    \field{journaltitle}{Mol. Phys.}
    \field{year}{2003}
  \endentry

  \entry{prascher2011gaussian}{article}{}
    \name{author}{5}{}{%
      {{hash=PBP}{%
         family={Prascher},
         familyi={P\bibinitperiod},
         given={Brian\bibnamedelima P},
         giveni={B\bibinitperiod\bibinitdelim P\bibinitperiod},
      }}%
      {{hash=WDE}{%
         family={Woon},
         familyi={W\bibinitperiod},
         given={David\bibnamedelima E},
         giveni={D\bibinitperiod\bibinitdelim E\bibinitperiod},
      }}%
      {{hash=PKA}{%
         family={Peterson},
         familyi={P\bibinitperiod},
         given={Kirk\bibnamedelima A},
         giveni={K\bibinitperiod\bibinitdelim A\bibinitperiod},
      }}%
      {{hash=DJTH}{%
         family={Dunning\bibnamedelima Jr},
         familyi={D\bibinitperiod\bibinitdelim J\bibinitperiod},
         given={Thom\bibnamedelima H},
         giveni={T\bibinitperiod\bibinitdelim H\bibinitperiod},
      }}%
      {{hash=WAK}{%
         family={Wilson},
         familyi={W\bibinitperiod},
         given={Angela\bibnamedelima K},
         giveni={A\bibinitperiod\bibinitdelim K\bibinitperiod},
      }}%
    }
    \list{publisher}{1}{%
      {Springer}%
    }
    \strng{namehash}{PBP+1}
    \strng{fullhash}{PBPWDEPKADJTHWAK1}
    \field{labelnamesource}{author}
    \field{labeltitlesource}{title}
    \verb{doi}
    \verb 10.1007/s00214-010-0764-0
    \endverb
    \field{number}{1}
    \field{pages}{69\bibrangedash 82}
    \field{title}{Gaussian basis sets for use in correlated molecular
  calculations. VII. Valence, core-valence, and scalar relativistic basis sets
  for Li, Be, Na, and Mg}
    \field{volume}{128}
    \field{journaltitle}{Theor. Chem. Acc.}
    \field{year}{2011}
  \endentry

  \entry{helgaker_estrap_1997}{article}{}
    \name{author}{4}{}{%
      {{hash=HT}{%
         family={Helgaker},
         familyi={H\bibinitperiod},
         given={Trygve},
         giveni={T\bibinitperiod},
      }}%
      {{hash=KW}{%
         family={Klopper},
         familyi={K\bibinitperiod},
         given={Wim},
         giveni={W\bibinitperiod},
      }}%
      {{hash=KH}{%
         family={Koch},
         familyi={K\bibinitperiod},
         given={Henrik},
         giveni={H\bibinitperiod},
      }}%
      {{hash=NJ}{%
         family={Noga},
         familyi={N\bibinitperiod},
         given={Jozef},
         giveni={J\bibinitperiod},
      }}%
    }
    \strng{namehash}{HT+1}
    \strng{fullhash}{HTKWKHNJ1}
    \field{labelnamesource}{author}
    \field{labeltitlesource}{title}
    \verb{doi}
    \verb 10.1063/1.473863
    \endverb
    \verb{eprint}
    \verb https://doi.org/10.1063/1.473863
    \endverb
    \field{number}{23}
    \field{pages}{9639\bibrangedash 9646}
    \field{title}{Basis-set convergence of correlated calculations on water}
    \verb{url}
    \verb https://doi.org/10.1063/1.473863
    \endverb
    \field{volume}{106}
    \field{journaltitle}{J. Chem. Phys.}
    \field{year}{1997}
  \endentry

  \entry{HALKIER1998243}{article}{}
    \name{author}{7}{}{%
      {{hash=HA}{%
         family={Halkier},
         familyi={H\bibinitperiod},
         given={Asger},
         giveni={A\bibinitperiod},
      }}%
      {{hash=HT}{%
         family={Helgaker},
         familyi={H\bibinitperiod},
         given={Trygve},
         giveni={T\bibinitperiod},
      }}%
      {{hash=JP}{%
         family={Jørgensen},
         familyi={J\bibinitperiod},
         given={Poul},
         giveni={P\bibinitperiod},
      }}%
      {{hash=KW}{%
         family={Klopper},
         familyi={K\bibinitperiod},
         given={Wim},
         giveni={W\bibinitperiod},
      }}%
      {{hash=KH}{%
         family={Koch},
         familyi={K\bibinitperiod},
         given={Henrik},
         giveni={H\bibinitperiod},
      }}%
      {{hash=OJ}{%
         family={Olsen},
         familyi={O\bibinitperiod},
         given={Jeppe},
         giveni={J\bibinitperiod},
      }}%
      {{hash=WAK}{%
         family={Wilson},
         familyi={W\bibinitperiod},
         given={Angela\bibnamedelima K.},
         giveni={A\bibinitperiod\bibinitdelim K\bibinitperiod},
      }}%
    }
    \strng{namehash}{HA+1}
    \strng{fullhash}{HAHTJPKWKHOJWAK1}
    \field{labelnamesource}{author}
    \field{labeltitlesource}{title}
    \verb{doi}
    \verb https://doi.org/10.1016/S0009-2614(98)00111-0
    \endverb
    \field{issn}{0009-2614}
    \field{number}{3}
    \field{pages}{243 \bibrangedash  252}
    \field{title}{Basis-set convergence in correlated calculations on Ne, N2,
  and H2O}
    \verb{url}
    \verb http://www.sciencedirect.com/science/article/pii/S0009261498001110
    \endverb
    \field{volume}{286}
    \field{journaltitle}{Chem. Phys. Lett.}
    \field{year}{1998}
  \endentry

  \entry{kutzelnigg1992rates}{article}{}
    \name{author}{2}{}{%
      {{hash=KW}{%
         family={Kutzelnigg},
         familyi={K\bibinitperiod},
         given={Werner},
         giveni={W\bibinitperiod},
      }}%
      {{hash=MIJD}{%
         family={Morgan\bibnamedelima III},
         familyi={M\bibinitperiod\bibinitdelim I\bibinitperiod},
         given={John\bibnamedelima D},
         giveni={J\bibinitperiod\bibinitdelim D\bibinitperiod},
      }}%
    }
    \list{publisher}{1}{%
      {AIP Publishing}%
    }
    \strng{namehash}{KWMIJD1}
    \strng{fullhash}{KWMIJD1}
    \field{labelnamesource}{author}
    \field{labeltitlesource}{title}
    \verb{doi}
    \verb 10.1063/1.462811
    \endverb
    \field{number}{6}
    \field{pages}{4484\bibrangedash 4508}
    \field{title}{Rates of convergence of the partial-wave expansions of atomic
  correlation energies}
    \field{volume}{96}
    \field{journaltitle}{J. Chem. Phys.}
    \field{year}{1992}
  \endentry

  \entry{klopper2001highly}{article}{}
    \name{author}{1}{}{%
      {{hash=KW}{%
         family={Klopper},
         familyi={K\bibinitperiod},
         given={WIM},
         giveni={W\bibinitperiod},
      }}%
    }
    \list{publisher}{1}{%
      {Taylor \& Francis}%
    }
    \strng{namehash}{KW1}
    \strng{fullhash}{KW1}
    \field{labelnamesource}{author}
    \field{labeltitlesource}{title}
    \verb{doi}
    \verb 10.1080/00268970010017315
    \endverb
    \field{number}{6}
    \field{pages}{481\bibrangedash 507}
    \field{title}{Highly accurate coupled-cluster singlet and triplet pair
  energies from explicitly correlated calculations in comparison with
  extrapolation techniques}
    \field{volume}{99}
    \field{journaltitle}{Mol. Phys.}
    \field{year}{2001}
  \endentry

  \entry{weigend2002fully}{article}{}
    \name{author}{1}{}{%
      {{hash=WF}{%
         family={Weigend},
         familyi={W\bibinitperiod},
         given={Florian},
         giveni={F\bibinitperiod},
      }}%
    }
    \list{publisher}{1}{%
      {Royal Society of Chemistry}%
    }
    \strng{namehash}{WF1}
    \strng{fullhash}{WF1}
    \field{labelnamesource}{author}
    \field{labeltitlesource}{title}
    \verb{doi}
    \verb 10.1039/B204199P
    \endverb
    \field{number}{18}
    \field{pages}{4285\bibrangedash 4291}
    \field{title}{A fully direct RI-HF algorithm: Implementation, optimised
  auxiliary basis sets, demonstration of accuracy and efficiency}
    \field{volume}{4}
    \field{journaltitle}{Phys. Chem. Chem. Phys.}
    \field{year}{2002}
  \endentry

  \entry{vahtras1993integral}{article}{}
    \name{author}{3}{}{%
      {{hash=VO}{%
         family={Vahtras},
         familyi={V\bibinitperiod},
         given={O},
         giveni={O},
      }}%
      {{hash=AJ}{%
         family={Alml{\"o}f},
         familyi={A\bibinitperiod},
         given={J},
         giveni={J},
      }}%
      {{hash=FM}{%
         family={Feyereisen},
         familyi={F\bibinitperiod},
         given={MW},
         giveni={M\bibinitperiod},
      }}%
    }
    \list{publisher}{1}{%
      {Elsevier}%
    }
    \strng{namehash}{VO+1}
    \strng{fullhash}{VOAJFM1}
    \field{labelnamesource}{author}
    \field{labeltitlesource}{title}
    \verb{doi}
    \verb 10.1016/0009-2614(93)89151-7
    \endverb
    \field{number}{5}
    \field{pages}{514\bibrangedash 518}
    \field{title}{Integral approximations for LCAO-SCF calculations}
    \field{volume}{213}
    \field{journaltitle}{Chem. Phys. Lett.}
    \field{year}{1993}
  \endentry

  \entry{turney2012psi4}{article}{}
    \true{moreauthor}
    \name{author}{10}{}{%
      {{hash=TJM}{%
         family={Turney},
         familyi={T\bibinitperiod},
         given={Justin\bibnamedelima M},
         giveni={J\bibinitperiod\bibinitdelim M\bibinitperiod},
      }}%
      {{hash=SAC}{%
         family={Simmonett},
         familyi={S\bibinitperiod},
         given={Andrew\bibnamedelima C},
         giveni={A\bibinitperiod\bibinitdelim C\bibinitperiod},
      }}%
      {{hash=PRM}{%
         family={Parrish},
         familyi={P\bibinitperiod},
         given={Robert\bibnamedelima M},
         giveni={R\bibinitperiod\bibinitdelim M\bibinitperiod},
      }}%
      {{hash=HEG}{%
         family={Hohenstein},
         familyi={H\bibinitperiod},
         given={Edward\bibnamedelima G},
         giveni={E\bibinitperiod\bibinitdelim G\bibinitperiod},
      }}%
      {{hash=EFA}{%
         family={Evangelista},
         familyi={E\bibinitperiod},
         given={Francesco\bibnamedelima A},
         giveni={F\bibinitperiod\bibinitdelim A\bibinitperiod},
      }}%
      {{hash=FJT}{%
         family={Fermann},
         familyi={F\bibinitperiod},
         given={Justin\bibnamedelima T},
         giveni={J\bibinitperiod\bibinitdelim T\bibinitperiod},
      }}%
      {{hash=MBJ}{%
         family={Mintz},
         familyi={M\bibinitperiod},
         given={Benjamin\bibnamedelima J},
         giveni={B\bibinitperiod\bibinitdelim J\bibinitperiod},
      }}%
      {{hash=BLA}{%
         family={Burns},
         familyi={B\bibinitperiod},
         given={Lori\bibnamedelima A},
         giveni={L\bibinitperiod\bibinitdelim A\bibinitperiod},
      }}%
      {{hash=WJJ}{%
         family={Wilke},
         familyi={W\bibinitperiod},
         given={Jeremiah\bibnamedelima J},
         giveni={J\bibinitperiod\bibinitdelim J\bibinitperiod},
      }}%
      {{hash=AML}{%
         family={Abrams},
         familyi={A\bibinitperiod},
         given={Micah\bibnamedelima L},
         giveni={M\bibinitperiod\bibinitdelim L\bibinitperiod},
      }}%
    }
    \list{publisher}{1}{%
      {Wiley Online Library}%
    }
    \strng{namehash}{TJM+1}
    \strng{fullhash}{TJMSACPRMHEGEFAFJTMBJBLAWJJAML+1}
    \field{labelnamesource}{author}
    \field{labeltitlesource}{title}
    \verb{doi}
    \verb 10.1002/wcms.93
    \endverb
    \field{number}{4}
    \field{pages}{556\bibrangedash 565}
    \field{title}{Psi4: an open-source ab initio electronic structure program}
    \field{volume}{2}
    \field{journaltitle}{WIREs Comput. Mol. Sci.}
    \field{year}{2012}
  \endentry

  \entry{orca_4}{article}{}
    \name{author}{1}{}{%
      {{hash=NF}{%
         family={Neese},
         familyi={N\bibinitperiod},
         given={Frank},
         giveni={F\bibinitperiod},
      }}%
    }
    \strng{namehash}{NF1}
    \strng{fullhash}{NF1}
    \field{labelnamesource}{author}
    \field{labeltitlesource}{title}
    \verb{doi}
    \verb 10.1002/wcms.1327
    \endverb
    \verb{eprint}
    \verb https://onlinelibrary.wiley.com/doi/pdf/10.1002/wcms.1327
    \endverb
    \field{number}{1}
    \field{pages}{e1327}
    \field{title}{Software update: the ORCA program system, version 4.0}
    \verb{url}
    \verb https://onlinelibrary.wiley.com/doi/abs/10.1002/wcms.1327
    \endverb
    \field{volume}{8}
    \field{journaltitle}{WIREs Comput. Mol. Sci.}
    \field{year}{2018}
  \endentry

  \entry{aux_c_hidrogeno}{article}{}
    \name{author}{3}{}{%
      {{hash=WF}{%
         family={Weigend},
         familyi={W\bibinitperiod},
         given={Florian},
         giveni={F\bibinitperiod},
      }}%
      {{hash=KA}{%
         family={Köhn},
         familyi={K\bibinitperiod},
         given={Andreas},
         giveni={A\bibinitperiod},
      }}%
      {{hash=HC}{%
         family={Hättig},
         familyi={H\bibinitperiod},
         given={Christof},
         giveni={C\bibinitperiod},
      }}%
    }
    \strng{namehash}{WF+1}
    \strng{fullhash}{WFKAHC1}
    \field{labelnamesource}{author}
    \field{labeltitlesource}{title}
    \verb{doi}
    \verb 10.1063/1.1445115
    \endverb
    \verb{eprint}
    \verb https://doi.org/10.1063/1.1445115
    \endverb
    \field{number}{8}
    \field{pages}{3175\bibrangedash 3183}
    \field{title}{Efficient use of the correlation consistent basis sets in
  resolution of the identity MP2 calculations}
    \verb{url}
    \verb https://doi.org/10.1063/1.1445115
    \endverb
    \field{volume}{116}
    \field{journaltitle}{J. Chem. Phys.}
    \field{year}{2002}
  \endentry

  \entry{aux_c_otros}{article}{}
    \name{author}{1}{}{%
      {{hash=HC}{%
         family={Hättig},
         familyi={H\bibinitperiod},
         given={Christof},
         giveni={C\bibinitperiod},
      }}%
    }
    \list{publisher}{1}{%
      {The Royal Society of Chemistry}%
    }
    \strng{namehash}{HC1}
    \strng{fullhash}{HC1}
    \field{labelnamesource}{author}
    \field{labeltitlesource}{title}
    \verb{doi}
    \verb 10.1039/B415208E
    \endverb
    \field{issue}{1}
    \field{pages}{59\bibrangedash 66}
    \field{title}{Optimization of auxiliary basis sets for RI-MP2 and RI-CC2
  calculations: Core–valence and quintuple-\textzeta basis sets for H to Ar
  and QZVPP basis sets for Li to Kr}
    \verb{url}
    \verb http://dx.doi.org/10.1039/B415208E
    \endverb
    \field{volume}{7}
    \field{journaltitle}{Phys. Chem. Chem. Phys.}
    \field{year}{2005}
  \endentry

  \entry{autoaux}{article}{}
    \name{author}{3}{}{%
      {{hash=SGL}{%
         family={Stoychev},
         familyi={S\bibinitperiod},
         given={Georgi\bibnamedelima L.},
         giveni={G\bibinitperiod\bibinitdelim L\bibinitperiod},
      }}%
      {{hash=AAA}{%
         family={Auer},
         familyi={A\bibinitperiod},
         given={Alexander\bibnamedelima A.},
         giveni={A\bibinitperiod\bibinitdelim A\bibinitperiod},
      }}%
      {{hash=NF}{%
         family={Neese},
         familyi={N\bibinitperiod},
         given={Frank},
         giveni={F\bibinitperiod},
      }}%
    }
    \strng{namehash}{SGL+1}
    \strng{fullhash}{SGLAAANF1}
    \field{labelnamesource}{author}
    \field{labeltitlesource}{title}
    \verb{doi}
    \verb 10.1021/acs.jctc.6b01041
    \endverb
    \verb{eprint}
    \verb https://doi.org/10.1021/acs.jctc.6b01041
    \endverb
    \field{note}{PMID: 28005364}
    \field{number}{2}
    \field{pages}{554\bibrangedash 562}
    \field{title}{Automatic Generation of Auxiliary Basis Sets}
    \verb{url}
    \verb https://doi.org/10.1021/acs.jctc.6b01041
    \endverb
    \field{volume}{13}
    \field{journaltitle}{J. Chem. Theory Comput.}
    \field{year}{2017}
  \endentry

  \entry{efron84tibshirani}{book}{}
    \name{author}{2}{}{%
      {{hash=EB}{%
         family={Efron},
         familyi={E\bibinitperiod},
         given={B.},
         giveni={B\bibinitperiod},
      }}%
      {{hash=TR}{%
         family={Tibshirani},
         familyi={T\bibinitperiod},
         given={R.},
         giveni={R\bibinitperiod},
      }}%
    }
    \strng{namehash}{EBTR1}
    \strng{fullhash}{EBTR1}
    \field{labelnamesource}{author}
    \field{labeltitlesource}{title}
  \field{isbn}{\href{https://isbnsearch.org/isbn/9780412042317}{9780412042317}}
    \field{title}{An Introduction to the Bootstrap}
    \field{volume}{1}
    \field{journaltitle}{Monographs on Statistics and Applied Probability.
  Chapman \& Hall/CRC}
    \field{year}{84}
  \endentry

  \entry{Duminda_cv_benchmark}{article}{}
    \name{author}{3}{}{%
      {{hash=RDS}{%
         family={Ranasinghe},
         familyi={R\bibinitperiod},
         given={Duminda\bibnamedelima S.},
         giveni={D\bibinitperiod\bibinitdelim S\bibinitperiod},
      }}%
      {{hash=FMJ}{%
         family={Frisch},
         familyi={F\bibinitperiod},
         given={Michael\bibnamedelima J.},
         giveni={M\bibinitperiod\bibinitdelim J\bibinitperiod},
      }}%
      {{hash=PGA}{%
         family={Petersson},
         familyi={P\bibinitperiod},
         given={George\bibnamedelima A.},
         giveni={G\bibinitperiod\bibinitdelim A\bibinitperiod},
      }}%
    }
    \strng{namehash}{RDS+1}
    \strng{fullhash}{RDSFMJPGA1}
    \field{labelnamesource}{author}
    \field{labeltitlesource}{title}
    \verb{doi}
    \verb 10.1063/1.4935972
    \endverb
    \verb{eprint}
    \verb https://doi.org/10.1063/1.4935972
    \endverb
    \field{number}{21}
    \field{pages}{214110}
    \field{title}{Core-core and core-valence correlation energy atomic and
  molecular benchmarks for Li through Ar}
    \verb{url}
    \verb https://doi.org/10.1063/1.4935972
    \endverb
    \field{volume}{143}
    \field{journaltitle}{J. Chem. Phys.}
    \field{year}{2015}
  \endentry

  \entry{Peterson2012}{article}{}
    \name{author}{3}{}{%
      {{hash=PKA}{%
         family={Peterson},
         familyi={P\bibinitperiod},
         given={Kirk\bibnamedelima A.},
         giveni={K\bibinitperiod\bibinitdelim A\bibinitperiod},
      }}%
      {{hash=FD}{%
         family={Feller},
         familyi={F\bibinitperiod},
         given={David},
         giveni={D\bibinitperiod},
      }}%
      {{hash=DDA}{%
         family={Dixon},
         familyi={D\bibinitperiod},
         given={David\bibnamedelima A.},
         giveni={D\bibinitperiod\bibinitdelim A\bibinitperiod},
      }}%
    }
    \strng{namehash}{PKA+1}
    \strng{fullhash}{PKAFDDDA1}
    \field{labelnamesource}{author}
    \field{labeltitlesource}{title}
    \verb{doi}
    \verb 10.1007/s00214-011-1079-5
    \endverb
    \field{issn}{1432-2234}
    \field{number}{1}
    \field{pages}{1079}
    \field{title}{Chemical accuracy in ab initio thermochemistry and
  spectroscopy: current strategies and future challenges}
    \verb{url}
    \verb https://doi.org/10.1007/s00214-011-1079-5
    \endverb
    \field{volume}{131}
    \field{journaltitle}{Theor. Chem. Acc.}
    \field{year}{2012}
    \warn{\item Invalid format of field 'month'}
  \endentry

  \entry{Feller_ccsdt_convergence}{article}{}
    \name{author}{3}{}{%
      {{hash=FD}{%
         family={Feller},
         familyi={F\bibinitperiod},
         given={David},
         giveni={D\bibinitperiod},
      }}%
      {{hash=PKA}{%
         family={Peterson},
         familyi={P\bibinitperiod},
         given={Kirk\bibnamedelima A.},
         giveni={K\bibinitperiod\bibinitdelim A\bibinitperiod},
      }}%
      {{hash=GHJ}{%
         family={Grant\bibnamedelima Hill},
         familyi={G\bibinitperiod\bibinitdelim H\bibinitperiod},
         given={J.},
         giveni={J\bibinitperiod},
      }}%
    }
    \strng{namehash}{FD+1}
    \strng{fullhash}{FDPKAGHJ1}
    \field{labelnamesource}{author}
    \field{labeltitlesource}{title}
    \verb{doi}
    \verb 10.1063/1.3613639
    \endverb
    \verb{eprint}
    \verb https://doi.org/10.1063/1.3613639
    \endverb
    \field{number}{4}
    \field{pages}{044102}
    \field{title}{On the effectiveness of CCSD(T) complete basis set
  extrapolations for atomization energies}
    \verb{url}
    \verb https://doi.org/10.1063/1.3613639
    \endverb
    \field{volume}{135}
    \field{journaltitle}{J. Chem. Phys.}
    \field{year}{2011}
  \endentry

  \entry{core_dft}{article}{}
    \name{author}{3}{}{%
      {{hash=RDS}{%
         family={Ranasinghe},
         familyi={R\bibinitperiod},
         given={Duminda\bibnamedelima S.},
         giveni={D\bibinitperiod\bibinitdelim S\bibinitperiod},
      }}%
      {{hash=FMJ}{%
         family={Frisch},
         familyi={F\bibinitperiod},
         given={Michael\bibnamedelima J.},
         giveni={M\bibinitperiod\bibinitdelim J\bibinitperiod},
      }}%
      {{hash=PGA}{%
         family={Petersson},
         familyi={P\bibinitperiod},
         given={George\bibnamedelima A.},
         giveni={G\bibinitperiod\bibinitdelim A\bibinitperiod},
      }}%
    }
    \strng{namehash}{RDS+1}
    \strng{fullhash}{RDSFMJPGA1}
    \field{labelnamesource}{author}
    \field{labeltitlesource}{title}
    \verb{doi}
    \verb 10.1063/1.4935973
    \endverb
    \verb{eprint}
    \verb https://doi.org/10.1063/1.4935973
    \endverb
    \field{number}{21}
    \field{pages}{214111}
    \field{title}{A density functional for core-valence correlation energy}
    \verb{url}
    \verb https://doi.org/10.1063/1.4935973
    \endverb
    \field{volume}{143}
    \field{journaltitle}{J. Chem. Phys.}
    \field{year}{2015}
  \endentry

  \entry{mp4}{article}{}
    \name{author}{2}{}{%
      {{hash=KR}{%
         family={Krishnan},
         familyi={K\bibinitperiod},
         given={R.},
         giveni={R\bibinitperiod},
      }}%
      {{hash=PJA}{%
         family={Pople},
         familyi={P\bibinitperiod},
         given={J.\bibnamedelima A.},
         giveni={J\bibinitperiod\bibinitdelim A\bibinitperiod},
      }}%
    }
    \strng{namehash}{KRPJA1}
    \strng{fullhash}{KRPJA1}
    \field{labelnamesource}{author}
    \field{labeltitlesource}{title}
    \field{abstract}{%
    Abstract An approximate fourth-order expression for the electron
  correlation energy in the Møller–Plesset perturbation scheme is proposed.
  It takes into account all the contributions to the fourthorder energy
  neglecting only those of the triple-substituted determinants. It is size
  consistent and correct to fourth order for an assembly of isolated
  two-electron systems. Illustrative calculations are reported for a series of
  small molecules.%
    }
    \verb{doi}
    \verb 10.1002/qua.560140109
    \endverb
    \verb{eprint}
    \verb https://onlinelibrary.wiley.com/doi/pdf/10.1002/qua.560140109
    \endverb
    \field{number}{1}
    \field{pages}{91\bibrangedash 100}
    \field{title}{Approximate fourth-order perturbation theory of the electron
  correlation energy}
    \verb{url}
    \verb https://onlinelibrary.wiley.com/doi/abs/10.1002/qua.560140109
    \endverb
    \field{volume}{14}
    \field{journaltitle}{Int. J. Quantum Chem.}
    \field{year}{1978}
  \endentry

  \entry{kendall1992a}{article}{}
    \name{author}{3}{}{%
      {{hash=KRA}{%
         family={Kendall},
         familyi={K\bibinitperiod},
         given={Rick\bibnamedelima A.},
         giveni={R\bibinitperiod\bibinitdelim A\bibinitperiod},
      }}%
      {{hash=DTH}{%
         family={Dunning},
         familyi={D\bibinitperiod},
         given={Thom\bibnamedelima H.},
         giveni={T\bibinitperiod\bibinitdelim H\bibinitperiod},
      }}%
      {{hash=HRJ}{%
         family={Harrison},
         familyi={H\bibinitperiod},
         given={Robert\bibnamedelima J.},
         giveni={R\bibinitperiod\bibinitdelim J\bibinitperiod},
      }}%
    }
    \strng{namehash}{KRA+1}
    \strng{fullhash}{KRADTHHRJ1}
    \field{labelnamesource}{author}
    \field{labeltitlesource}{title}
    \verb{doi}
    \verb 10.1063/1.462569
    \endverb
    \field{title}{Electron affinities of the first-row atoms revisited.
  Systematic basis sets and wave functions}
    \field{volume}{96}
    \field{journaltitle}{J. Chem. Phys.}
    \field{year}{1992}
  \endentry
\endsortlist

	\blx@bblend
	\endgroup
	\csnumgdef{blx@labelnumber@\the\c@refsection}{0}%
	\iftoggle{blx@reencode}{\blx@reencode}{}}

\usepackage{ragged2e}
\usepackage{booktabs}
\usepackage{tabularx}
\newcommand{\ra}[1]{\renewcommand{\arraystretch}{#1}}
\usepackage{placeins}

\usepackage[misc]{ifsym}

\usepackage{float}
\restylefloat{table}

\usepackage{abstract}
\begin{document}

	\title{\Large\textsf{\textbf{\MakeUppercase{{Prediction of the inner-shell contribution to the correlation energy through DLPNO-CEPA/1 and Scaled same-spin second order Møller–Plesset perturbation theory}}}}}
	\author{\small \textsf{\scshape{Hernán R. Sánchez}}}	
	\date{}
	
	\twocolumn[
	\maketitle
	\begin{center}
	{\vspace{-0.5cm}\small Centro de Química Inorgánica. CONICET  La Plata - UNLP, Argentina.	
		
		\footnotesize {\scriptsize\Letter} \textsf{\href{mailto:hernan.sanchez@quimica.unlp.edu.ar}{hernan.sanchez@quimica.unlp.edu.ar}}}
\end{center}
	\begin{onecolabstract}
 \vspace{-0.2cm}The use of two low cost methods for the prediction of the inner-shells contribution to the correlation energy is analyzed. The Spin-Component-Scaled second order Møller–Plesset perturbation theory (SCS-MP2) was reparameterized for the prediction of such contributions.  The best results are found when only the same spin term is considered (SSS-MP2).   The Coupled Electron Pair Approximation (CEPA) using the domain based local pair natural orbital approximation (DLPNO) was also studied for the same purpose.   The methods were tested on the W4-11 test set using basis sets up to quadruple zeta quality. The SSS-MP2 proved to be a marked improvement upon MP2 decreasing the root-mean-square-error (RMSE) from 0.443 to 0.302 kcal mol$^{-1}$.  The RMSE of DLPNO-CEPA/1 in the test set is only 0.147 kcal mol$^{-1}$ and its computational cost is very low considering the  intended applications.  Furthermore, a linear combination of both methods decreased the RMSE to 0.118 kcal mol$^{-1}$.
 
  		\vspace*{0.8cm}
	\end{onecolabstract}	
	]

\section{Introduction}

Most \textit{ab initio} post Hartree-Fock calculations are performed within the frozen core approximation (FC) in which the lowest-lying molecular orbitals are constrained to remain fully-occupied. This helps to avoid considerable computational cost, and is justified by the fact that inner-shell electrons are less sensitive to the molecular environment than the valence electrons, so their effects are similar in reactants and products leading to large compensations in reaction energies. This contribution to the correlation energy must be taken into account in highly accuracy calculations, especially for atomization reactions because the compensation is smaller and  contributions of around 2 kcal mol$^{-1}$ are not uncommon even for small molecules. Inner-shell contribution to reaction energy is becoming increasingly important as the accuracy requirements of many current calculations  are reaching the magnitude of such contributions.

Accurate non-frozen core (AE) calculations are normally forbidden by computational cost because they require to consider larger number of configurations and using larger one-electron basis sets. Inner-shell contribution may be accounted through the addition supposition\supercite{feller2008survey} as the difference between the energy of an AE calculation and a FC one, employing the same often smaller yet large basis set. Obtaining accurate results though this procedure is still very resource and time intensive, and it is often performed at levels of theory similar to coupled cluster simples and doubles with perturbational triples contribution (CCSD(T))\supercite{RAGHAVACHARI1989479,ccsd_t_original} employing at least aug-cc-pwCVQZ basis set\supercite{karton2011w4,feller2015improved,feller2008survey}.  It has been pointed out that methods in which connected triple excitations are neglected (as CCSD)  largely underestimate the core correlation contribution\supercite{martin2000thermochemical}.

The main goal of this work is to find computationally cheap methods of reasonable accuracy for computing the inner-shell contribution to the total energy. For this, variants of  second-order Møller–Plesset perturbation theory\supercite{moller1934note} (MP2) and the coupled electron pair approximation\supercite{cepa_ref0,cepa_ref1,cepa_ref2,cepa_ref3,cepa_ref4,cepa_ref5} (CEPA) are investigated. Previous research on the use of MP2 for this purpose was done\supercite{martin1999towards,martin2000thermochemical} and the authors found that MP2 systematically underestimates the CCSD(T) core correlation contributions for a small set of molecules. Nevertheless, the inner-shell contribution computed at MP2 is not useless and has been included in composite calculation methods\supercite{curtiss1998gaussian,curtiss2007gaussian}.

The MP2 allows useful \textit{ab initio} estimations of electron correlation effects. Its relatively low computational cost permits its application to large systems that cannot be treated with more accurate common \textit{ab initio} methods. Also, as it shares similarities in basis set convergence properties with more computationally demanding methods, it has been used in conjunction with extrapolation techniques in order to estimate complete basis set values for the latter. For some years there has been a renewed interest in MP2, efforts have been focused on obtaining more efficient implementations\supercite{doser2008tighter,steele2006dual,weigend1997ri,DLPNO-MP2-F12} and modifications for improving accuracy\supercite{sosdistance,o2regularisation,o2regularisationfix,regoomp22018,bsseatenuation,attenuatedTZ,oo-mp2,oo-sos-mp2,SCS-MP2mod}.

The MP2 correlation energy ($E_c$) can be expressed as the sum of the contributions of electron pairs with the same spin ($E_{ss}$) and opposite spin ($E_{os}$)\supercite{petersson1985complete}

\begin{equation*}
E_c = E_{ss} + E_{os}
\end{equation*}
\noindent
where if Mulliken's notation\supercite{szabo1996modern} is used for two-electron integrals

\begin{align*} 
E_{os} &=  - \sum _{abij} \frac{(ai|jb)^2}{\epsilon _a + \epsilon_b -\epsilon_i - \epsilon_j}\\ 
E_{ss} &=  - \sum _{abij} \frac{(ai|jb)\left[(ai|jb)-(aj|ib) \right]}{\epsilon _a + \epsilon_b -\epsilon_i - \epsilon_j}
\end{align*}

In the above equations, the contributions to $E_{os}$  ($E_{ss}$)  include excitation of two electrons with opposite (same) spin from the orbitals $i$ and $j$ to the orbitals $a$ and $b$.   Grimme proposed the application of separate scaling factors for the same spin (SS) and opposite spin (OS) terms,

\begin{equation*}
E_c = c_{ss} E_{ss} + c_{os} E_{os}
\end{equation*}

\noindent and fitted the coefficients to a set of reaction energies computed at the QCISD(T)\supercite{pople1987quadratic} level of theory, obtaining the pioneer Spin Component Scaled MP2 (SCS-MP2) whose coefficients are $c_{os}=6/5$ and $c_{ss}=1/3$\supercite{grimme2003improved}. Although it emerged as a semi-empirical method posterior research provided theoretical justification for using scaling parameters not equal to one\supercite{Szabados2006,fink2010}.

Since then, many variants have been proposed with specific targets \supercite{jung2004scaled,neese2009assessment,szabados2011,distasio2007optimized,hill2007spin,SCSILMP2,Gordon2014,grimme_review_2012}. The complete neglect of the SS energies ($c_{ss}=0$) alternative was developed, giving rise to Scaled Opposite-Spin method\supercite{jung2004scaled} (SOS-MP2), which was very attractive in that the  scaling with system size can be reduced from fifth order for MP2 to fourth order for this method. Both SCS-MP2 and SOS-MP2 improved upon MP2 in many tests, and their global quality was found to be similar. However, it was found that to play down the importance of the same spin  contribution causes adverse effects for the long range interaction. In fact, a SCS-MP2 variant which neglects the opposite spin contribution was proposed for studding long range interactions\supercite{hill2007spin}. Given that scaling the SS and OS terms was fruitful for many applications, it could be expected that properly scaling those terms leads to better results for the inner-shell correlation energy. In this work those scaling coefficients are estimated.

The methods based on CEPA are commonly  more accurate and more computationally demanding than those based in MP2.  They often give better results than CCSD at slightly lower cost making them an interesting alternative for our purpose.  It is noteworthy that complete basis set (CBS) CCSD does not provide very accurate results for the inner-shell contribution to the Total Atomization Energy (TAE) even after a global rescaling. This can be readily seen from the supplementary data of the reference \citenum{martin2019}. CEPA based methods can be applied under the domain based local pair natural orbital approximation (DLPNO) which drastically reduces the computational cost  with almost no loss of accuracy. It has been shown that this approximation gives excelent results for inner-shell correlation energies\supercite{dlpno-ccsdt-core}. Among the many traditional variants of CEPA the one referred as CEPA/1 seems to be the more accurate\supercite{AHLRICHS197931,WENNMOHS2008217}. In this study the use of DLPNO-CEPA/1 for the prediction of the inner-shell correlation energy is analyzed.

\section{Methods}
	
In order to parameterize the studied methods or to assess their accuracy a set with accurate reference values of the non-relativistic inner-shell correlation contribution to energy is needed. The W4-11 test set\supercite{karton2011w4} fulfills this requirement and was employed in the present work.  It contains 140 diverse chemical species whose TAE were calculated at very high level of theory. In this work the $\text{H}_2$ molecule was removed from the test set because it only has valence electrons. The radicals $\text{FO}_2$ and $\text{ClOO}$ were also removed due to some convergence problems.

For open-shell cases, the authors of the W4-11 test set chose the Restricted Open-shell Hartree–Fock\supercite{RevModPhys.32.179} (ROHF) reference, and used the Werner–Knowles–Hampel version of the restricted open-shell CCSD(T)\supercite{Knowles_ccsd_t,errata_knowles}. The same reference was used in this study because doing so is more advantageous in comparative terms. Core-core and core-valence contributions to the correlation energy were computed as the difference between full and frozen core calculations using the same basis set.

 In the case of MP2, the aug-cc-pwCVTZ and aug-cc-pwCVQZ\supercite{peterson2002accurate} basis sets were employed, except for hydrogen and beryllium for which the corresponding cc-pV$n$Z\supercite{dunning1989gaussian} and the Iron's version of the cc-pCV$n$Z\supercite{iron2003alkali,prascher2011gaussian} basis set, respectively,  were used.  The Helgaker's extrapolation scheme\supercite{helgaker_estrap_1997,HALKIER1998243} was employed for the OS contribution to the correlation energy, while for the SS contribution a basis set dependence of the form $E_{ss,z} = E_{ss,\infty}+A z^{-5}$ (where $A$ is a constant value and $z$ equals two for double-zeta sets, three for triple-zeta sets, etc.) was assumed\supercite{kutzelnigg1992rates,klopper2001highly}. The Resolution of the Identity (RI) approximation\supercite{weigend2002fully,weigend1997ri,vahtras1993integral} for the two-electron integrals was employed to speed up the computations. The mentioned calculations were performed using the Psi4 program\supercite{turney2012psi4}.

The DLPNO-CEPA/1 calculations were done with ORCA 4.2\supercite{orca_4}. The use of the DLPNO approximation requires the specification of many thresholds which control its accuracy. The default values were used. In accordance with the goal of finding a cheap method, the basis set used for DLPNO-CEPA/1 are smaller than those for MP2, as the former is substantially more computationally expensive. The  cc-pwCVTZ and cc-pwCVQZ\supercite{peterson2002accurate} basis sets were used in place of their augmented counterparts to make predictions. The aug-cc-pwCVTZ was also employed for comparative purposes. The The standard choice of the auxiliary basis sets\supercite{aux_c_hidrogeno,aux_c_otros} required by the DLPNO method were done. Hereinafter the shorthand (a)wCVnZ (n=T,Q) will be used for (aug-)cc-pwCVnZ, denoting by n=[T,Q] the triple and quadruple zetas extrapolation. Again, the Helgaker's  method was used for extrapolation to basis set limits. In the case of beryllium, the auxiliary basis set were automatically built by ORCA\supercite{autoaux} using the more conservative available parameters.

To assess the predictive power of a model it is required a test set independent of the training set used for the estimation of the model parameters. In this work, the models were validated through the Leave One Out (LOO) method\supercite{efron84tibshirani}. Having $n$ observations, it consist in perform $n$ parameter estimations using different subsets of $n-1$ observations, and in each case the remaining observation is used as the validation set. The $n$ residues thus obtained are used for the evaluation of the performance of the model when its parameters are fitted using $n-1$ observations. This procedure should be a slightly pessimistic estimation of the prediction power of the model parameterized using the $n$ observations.

\section{Results and Discussion}
The main results can be found in Table \ref{Table results}. The statistics used for quantifying  the predictive power were computed using the errors on the whole set of 137 chemical species for methods which does not include additional parameters, and using the LOO residues otherwise. The last row correspond to the errors due to  neglect of the inner-shell contribution in the test set.

As expected DLPNO-CEPA/1 is more accurate than MP2 and its derived methods. When energies computed with the former are scaled the TAE improve significantly for 3$\zeta$ and, to a lesser extent, for  4$\zeta$. The extrapolated wCV[T,Q]Z results appear to have reached values reasonably close to those of CBS and almost no improvement is obtained by simple scaling. This is hardly surprising given that CCSD(T)  provides almost converged values for basis sets of similar quality, see for example the references  \citenum{martin2019} and \citenum{Duminda_cv_benchmark}. It is noteworthy that no significant bias is found for DLPNO-CEPA/1/wCV[T,Q]Z, having a mean signed error (MSE) of only $-0.012$. Its mean absolute error (MAE) and the root-mean-square error  (RMSE) are 0.096 and 0.147 kcal mol$^{-1}$, respectively. This can be regarded as a success given the relatively low cost of the method.

In the case of MP2 large improvements are found by simple scaling even after performing the 3/4$\zeta$ extrapolation, which at the same times implies a large scaling coefficient ($\approx 1.5$). This suggests that MP2 inherently underestimate the inner-shell contribution. It should be noted that the basis sets used for MP2 are larger than those used for DLPNO-CEPA/1 because the former include diffuse functions. Their effect on the inner-shell correlation energy is small but still significant for few chemical species. For example, the RMSE of the comparison between DLPNO-CEPA/1 calculations using awCVTZ and wCVTZ is 0.110 kcal mol$^{-1}$ and reduces to 0.075 kcal mol$^{-1}$ if AlF$_3$ and SiF$_4$ are removed.

\begin{figure}[tbph!]
	\centering
	\includegraphics[width=1.\linewidth]{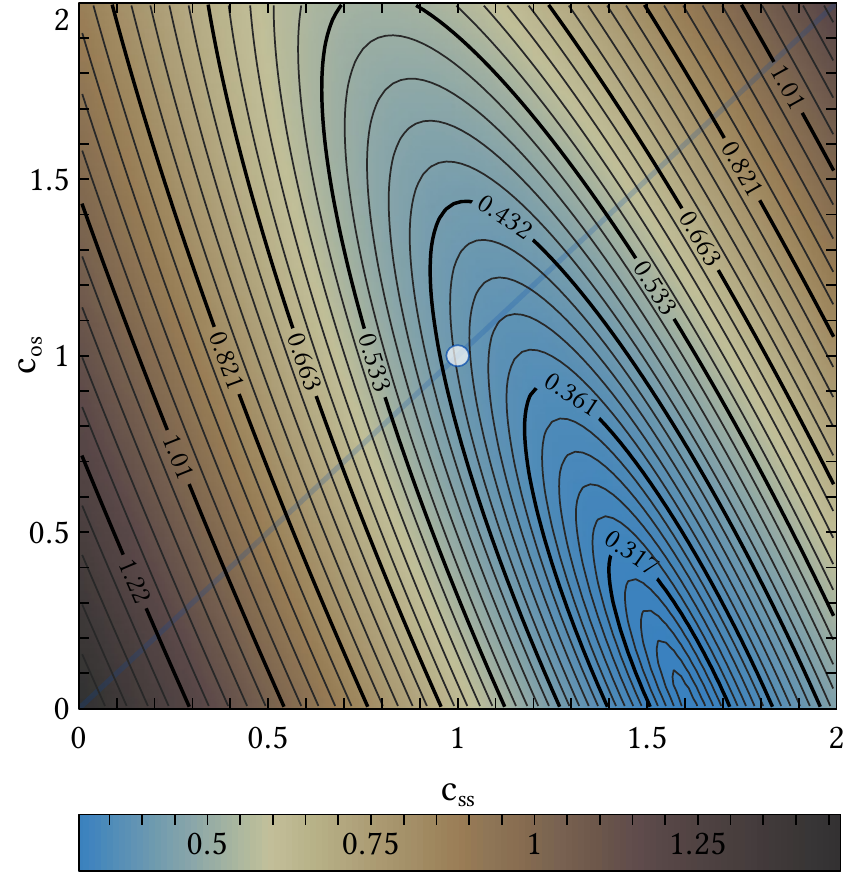}
	\caption{Influence of the scaling coefficients on the RMSE of the inner-shell contribution to the correlation energy using awCV[T,Q]Z.}
	\label{fig:graph01}
\end{figure}

The scaled same spin  MP2 (SSS-MP2) performs better than scaling MP2. That is, the inclusion of the opposite spin (OS) contribution employing the same scaling coefficient  degrades the MP2 performance. Separate scaling of both components gives the parameters $1.6289$ and $0.0094$ for the SS and OS contributions, respectively, and their corresponding standard deviation are 0.050 and 0.109. The almost null coefficient of the SO part indicates the convenience of neglecting the OS contribution. The effects of the scaling coefficients on the RMSE of the inner-shell contribution is represented in detail in Figure \ref{fig:graph01} and corroborate this finding.

\begin{figure}[tbph!]
	\centering
	\includegraphics[scale=1]{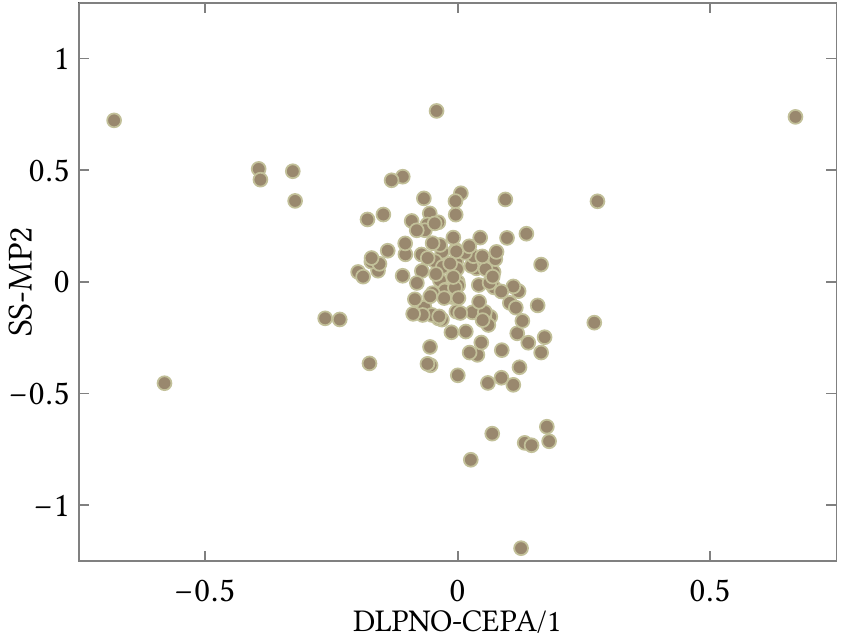}
	\caption{Residues of SSS-MP2 vs. DLPNO-CEPA/1 for wCV[T,Q]Z in units of kcal mol$^{-1}$.}
	\label{fig:graph02}
\end{figure}

In Figure \ref{fig:graph02} the residues corresponding to SSS-MP2 are plotted against those of DLPNO-CEPA/1 employing wCV[T,Q]Z in both cases. They seems to be uncorrelated as can be seen at a glance. This suggests that an interesting alternative could be improving upon DLPNO-CEPA/1/wCV[T,Q]Z by making a linear combination of this method and SSS-MP2/awCV[T,Q]Z. The results for such combination can be found in Table \ref{Table results}. The MAE and RMSE for this semi-empirical method are  0.076 and 0.118 kcal mol$^{-1}$, respectively. The median of the absolute values of the residues (Mdn(AE)) is only 0.050 kcal mol$^{-1}$. Those results are a significant improvement upon the DLPNO-CEPA/1/wCV[T,Q]Z. Nevertheless, the accuracy of both methods is enough for most practical applications. The inclusion of the OS was also considered but again this proves to be fruitless.

The most frequently used low cost alternative to CCSD(T) for computing the studied inner-shell contribution is MP2. A more detailed visual comparison of the latter with SSS-MP2 and DLPNO-CEPA/1 can be done through the histogram of Figure \ref{fig:graph03}. It is important to note that for MP2 and SSS-MP2 some values fall outside the plotted range.

\FloatBarrier
\begin{table*}[htb]\footnotesize\ra{1.3}
	\newcolumntype{Y}{>{\raggedleft\arraybackslash}X}
	\newcolumntype{Z}{>{\centering\arraybackslash}X}
	\begin{tabularx}{1.\textwidth}{lZZZZZZZZ}\arrayrulecolor{gray} \toprule  \arrayrulecolor{histobrown}
		&\multicolumn{1}{Z}{\scriptsize\bfseries\sffamily MAE} &\multicolumn{1}{Z}{\scriptsize\bfseries\sffamily RMSE}&\multicolumn{1}{Z}{\scriptsize\bfseries\sffamily STD}&\multicolumn{1}{Z}{\scriptsize\bfseries\sffamily LAE}&\multicolumn{1}{Z}{\scriptsize\bfseries\sffamily MSE }&\multicolumn{1}{Z}{\scriptsize\bfseries\sffamily Mdn(AE)}&\multicolumn{1}{Z}{\scriptsize\bfseries\sffamily $s_1$}&\multicolumn{1}{Z}{\scriptsize\bfseries\sffamily $s_2$}\\\midrule
		{\sffamily\scriptsize\textbf{{\color{histobrown}(a)wCVTZ}}$^{\text{\normalfont(a)}}$}&&&&&&&&\\
		\sffamily\scriptsize 	MP2 &0.483 &0.617&0.426&1.849&$-0.447$&0.447&-&-\\
		\sffamily\scriptsize 	DLPNO-CEPA/1 & 0.246&0.302&0.201&1.027&$-0.226$&0.224&-&-\\		
		\sffamily\scriptsize 	MP2 scaled & 0.347&0.543&0.479&2.449&$-0.258$&0.255&1.2881$\,\,$\textit{0.043}&-\\
		\sffamily\scriptsize 	SOS-MP2 & 0.733&0.923&0.717&2.949&$-0.584$&0.696&3.2026$\,\,$\textit{0.214}&\\
		\sffamily\scriptsize 	SSS-MP2 &0.262 &0.368&0.365&1.609&$-0.054$&0.181&1.6796$\,\,$\textit{0.037}&-\\
		\sffamily\scriptsize 	DLPNO-CEPA/1 scaled &0.153&0.239&0.221&1.495&$-0.094$&0.111&1.1490$\,\,$\textit{0.016}&-\\[8pt]
		{\sffamily\scriptsize\textbf{{\color{histobrown}(a)wCVQZ}}$^{\text{\normalfont(a)}}$}&&&&&&&&\\
		\sffamily\scriptsize MP2 &0.441 &0.561&0.372&1.818&$-0.422$&0.413&-&-\\
		\sffamily\scriptsize DLPNO-CEPA/1 &0.130 &0.176&0.143&0.820&$-0.102$&0.110&-&-\\		
		\sffamily\scriptsize MP2 scaled & 0.307&0.489&0.424&2.121&$-0.246$&0.226&1.2577$\,\,$\textit{0.038}&-\\
		\sffamily\scriptsize SOS-MP2 &0.697 &0.891&0.645&3.361&$-0.617$&0.653&3.0961$\,\,$\textit{0.196}&-\\
		\sffamily\scriptsize SSS-MP2 & 0.231&0.316&0.316&1.294&$-0.025$&0.162&1.6444$\,\,$\textit{0.031}&\\
		\sffamily\scriptsize DLPNO-CEPA/1 scaled &0.107 &0.163&0.155&0.972&$-0.052$&0.073&1.0497$\,\,$\textit{0.010}&-\\[8pt]
		{\sffamily\scriptsize\textbf{{\color{histobrown}(a)wCV[T,Q]Z}}$^{\text{\normalfont(a)}}$}&&&&&&&&\\
		\sffamily\scriptsize MP2 & 0.324&0.443&0.343&1.659&$-0.282$&0.274&-&-\\
		\sffamily\scriptsize DLPNO-CEPA/1 &0.096&0.147 &0.147&0.680&$-0.012$&0.062&-&-\\		
		\sffamily\scriptsize MP2 scaled &0.239 &0.407&0.375&1.785&$-0.162$&0.158&1.1463$\,\,$\textit{0.028}&-\\
		\sffamily\scriptsize SOS-MP2 & 0.691&0.903&0.620&3.534&$-0.658$&0.626&2.9019$\,\,$\textit{0.188}&-\\
		\sffamily\scriptsize SSS-MP2 &0.222 &0.302&0.303&1.199&$-0.017$&0.156&1.6324$\,\,$\textit{0.029}&-\\
		\sffamily\scriptsize DLPNO-CEPA/1 scaled &0.094&0.145&0.141 &0.680&$-0.033$&0.064&0.9812$\,\,$\textit{0.008}&-\\[4pt]
		\sffamily\scriptsize SSS-MP2 + SOS-MP2  &0.222 &0.302&0.303&1.199&$-0.017$&0.156&1.6289$\,\,$\textit{0.050}&0.0094$\,\,$\textit{0.109}\\
		\sffamily\scriptsize DLPNO-CEPA/1 + SSS-MP2  &0.076&0.118&0.118&0.694&$-0.017$&0.050&0.7522$\,\,$\textit{0.027}&0.3998$\,\,$\textit{0.046}\\\midrule
		\sffamily\scriptsize\bfseries Total inner-shell contrib. &1.135&1.468&0.962&3.760&1.112&0.865&-&-\\\arrayrulecolor{gray}\bottomrule
	\end{tabularx}
	\caption{Statistics of residues the of the studied methods.\label{Table results}}	
	$\,$\\
	
	$^{\text{\normalfont(a)}}$: Diffuse functions only for MP2 and derived methods. MAE: Mean absolute error. RMSE: Root mean square errors. STD: Standard deviation of the residues. LAE: Largest absolute value of residues. MSE: Mean signed error. Mdn(AE): Median of absolute errors. $s_1$: (1st) Scaling coefficient and its standard deviation (in italics). $s_2$: (2nd) Scaling coefficient and its standard deviation (in italics). See main text for details.
\end{table*}
\FloatBarrier

\begin{figure}[tbph!]
	\centering
	\includegraphics[width=1.\linewidth]{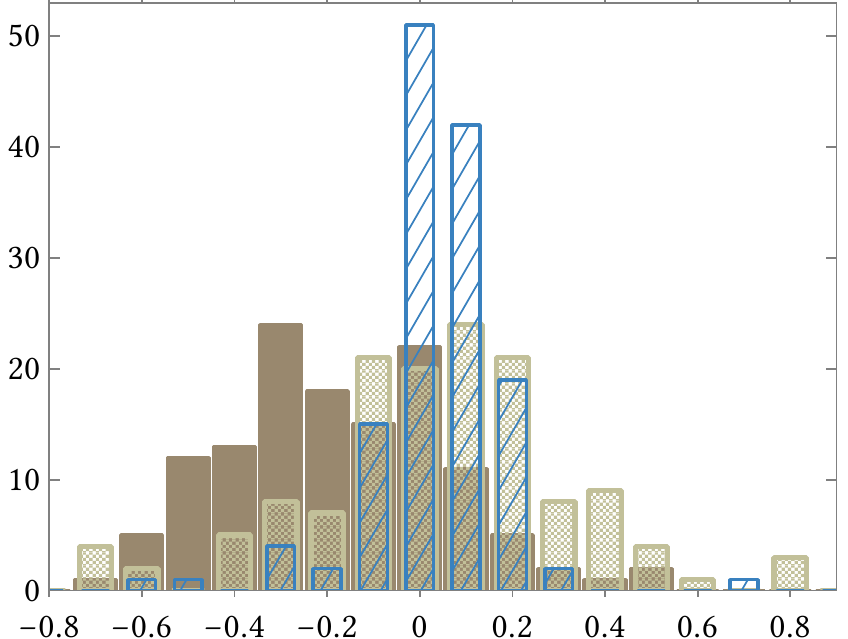}
	\caption{Section of the histogram of the residues of DLPNO-CEPA/1 ({\color{histoblue}$\blacksquare$}), MP2 ({\color{histobrown}$\blacksquare$}) and SS-MP2 ({\color{histoyellow}$\blacksquare$}) for awCV[T,Q]Z. }
	\label{fig:graph03}
\end{figure}

The success of Grimme's pioner method in which the OS component has a large scaling factor (6/5) turns the above results somewhat striking. Because of this, the SCS-MP2 method was parameterized with the observations of the present test set for valence only correlation in order to verify the correspondence with Grimme's results. The scaling factors found for the extrapolated awCV[T,Q]Z contributions  are 1.1455  and 0.0905 for the OS and SS  terms, respectively. Their corresponding standard deviation are 0.022 and 0.062. In the same order, for awCVTZ the obtained values are 1.2713, 0.0030, 0.023 and 0.061. Finally for awCVQZ:  1.1951, 0.0582, 0.020 and 0.056. These results are in accordance with those by Grimme.

\subsection{Comparison with other methods}	
From a practical standpoint, an useful comparison requires consideration of the accuracy needs. Normally the inner-shell contribution to the correlation energy is computed to be included in some composite method\supercite{Peterson2012}. Most approaches compute the valence correlation energy up to the gold standard method of quantum chemistry, i.e. CCSD(T). Reaching chemical accuracy for reaction energies (1 kcal mol$^{-1}$) requires computing the latter contribution by extrapolation using basis sets of triple and quadruple zeta quality, in the case of small molecules\supercite{Feller_ccsdt_convergence}.  This information is indicative as depends moderately on the extrapolation method employed and largely on the particular studied chemical specie, being the system size a major factor. The errors of the best method from this work are comparable with those of the  valence CCSD(T) contribution when the latter is obtained by extrapolation of quadruple/quintuple or quintuple/sextuple zeta quality basis sets\supercite{Feller_ccsdt_convergence}. In such cases, the computational cost of DLPNO-CEPA/1/wCVQZ is insignificant. The methods studied in this study become inadequate  if higher accuracy is desired, because the additional computational cost required (for computing higher excitations or reducing the  basis set incompleteness errors) justifies the computation of the core correlation contribution though CCSD(T) or similar methods.

It is worth mentioning the work of Ranasinghe and coworkers\supercite{core_dft}. They followed a density functional approach and obtained a cheap and accurate method for the prediction of the inner-shell correlation. The comparison between their results and those of this work has approximate character as the test sets employed are not the same. However, many conclusions can be drawn. In their test, their method CV-DFT/3Za1Pa has a RMSE of 0.27 kcal mol$^{-1}$ outperforming the raw values of MP2, MP4\supercite{mp4} and CCSD. The RMSE should not be directly compared between works because it tends to increase with the number of included observations and their test set is larger than the one used in this work. After correcting by sets cardinality their RMSE is 0.23 kcal mol$^{-1}$ approximately.   This suggests that the accuracy of CV-DFT/3Za1Pa (RMSE $\approx 0.23$) is in between SSS-MP2 (RMSE $\approx 0.30$ ) and DLPNO-CEPA/1 (RMSE $\approx 0.15$). The order of accuracy seems to be CCSD(T)$>$ DLPNO-CEPA/1 + SSS-MP2 $>$  DLPNO-CEPA/1 $>$ CV-DFT/3Za1Pa $>$ SSS-MP2. And raw CCSD, MP4 and MP2 falling behind. Although CV-DFT/3Za1Pa seems to behave slightly better than SSS-MP2, the natural choice would be SSS-MP2 because the computation of the SS contribution to the MP2 correlation energy is implemented in many publicly available software.

The scaling coefficients found for SS-MP2 are roughly 1.63, 1.64 and 1.68 for awCV[T,Q]Z, awCVQZ and awCVTZ, respectively. Employing an even smaller basis set should increase the scaling parameter. Grant Hill and Platts found good results neglecting the OS contribution and setting the SS scaling parameter to 1.75 in their spin-component  scaled  for  nucleo-bases variant (SCSN-MP2)\supercite{hill2007spin}. They used the aug-cc-pVTZ\supercite{kendall1992a,dunning1989gaussian}  basis set which is smaller than awCVTZ used in this study. Thus, the fitting process returned almost the same parameters in both works. Notably, they studied long range interactions while in the present work the focus is in those of short range. However, the valence only correlation contribution to the reaction energies associated mainly to covalent bonds is much better represented by the OS contribution, as mentioned before. Lochan and coworkers proposed a  distance-dependent scaling of the opposite spin correlation energy given rise to the MOS-MP2 method\supercite{sosdistance}. It would be worthwhile to use the results obtained in this study  as an aid for developing improved distance-dependent scaling methods.

\section{Summary and conclusions}	
Within the \textit{ab initio} framework the inner-shell contribution to the correlation energy has been commonly omitted for computational savings thanks to large error cancellations.  The current standards of accuracy are becoming this contribution more and more important. For reasons of efficiency composite methods are used and this contribution is normally computed at levels of theory lower than those used for the valence part. However, CCSD(T) is often used for all electron calculations which leads to very high computational cost. Another common alternative is to use MP2 which implies larger errors.

In this work low cost alternatives were studied on a subset of W4-11 containing 137 varied chemical species mainly as MP2 replacements. Using (a)wCV[T,Q]Z the RMSE of (SSS-MP2) DLPNO-CEPA/1 is (0.302) 0.147 kcal mol$^{-1}$. The corresponding value for  MP2  is 0.443  kcal mol$^{-1}$. The common context in which these calculations are performed suggest that DLPNO-CEPA/1/wCV[T,Q]Z are affordable and thus recommended. If lower cost methods are required, the use of SSS-MP2 in place of the original MP2 method is suggested. 

In the light of the findings of this study further research, for reviewing and expanding the knowledge regarding the underlying physics behind the scaling coefficients of the SCS-MP2, appears to be worthwhile.

\section{Acknowledgment}	

\textit{The author expresses his thanks to Vanesa González for her company and patience without which the completion of this and other leisure time research projects would not have been possible.}

\linespread{1.1} 
\printbibliography

\end{document}